\documentclass{aa}
\usepackage{graphicx}
\usepackage{color}
\usepackage{txfonts}
\begin{document}
    \title{Variable stars in the open cluster NGC~6791 \\ and its surrounding field}

   \author{
          F. De Marchi
          \inst{1},
          E. Poretti
          \inst{2},
          M. Montalto
          \inst{1},
          G. Piotto
          \inst{1},
          S. Desidera
          \inst{3},
          L.R. Bedin
          \inst{4},\\
          \vspace{0.1cm}
          R. Claudi
          \inst{3},
          A. Arellano Ferro
          \inst{5},
          H. Bruntt
          \inst{6},
          P.B. Stetson
          \inst{7}
          }

   \authorrunning{De Marchi et al.}

   \offprints{F.~De~Marchi\\
              \email{fabrizio.demarchi@unipd.it} }

   \institute{Dipartimento di Astronomia, Universit\`a di Padova,
              Vicolo dell'Osservatorio 2, I-35122, Padova, Italy
              \and
              INAF -- Osservatorio Astronomico di Brera,
              Via E. Bianchi 46, 23807 Merate (LC), Italy
              \and
              INAF -- Osservatorio Astronomico di Padova,
              Vicolo dell'Osservatorio 5, I-35122, Padova, Italy
              \and
              Space Telescope Science Institute, 3700 San Martin Drive, Baltimore, MD 21218, USA
              {\tt bedin@stsci.edu}
              \and
              Instituto de Astronom\'{\i}a, Universidad Nacional Aut\'{o}noma de M\'{e}xico, DF, Mexico
              \and
              School of Physics A28, University of Sydney, NSW 2006, Australia
              \and
              Herzberg Institute of Astrophysics, Victoria, Canada
             }

   \date{Received, accepted}

   \abstract
   {}
 {This work presents a high--precision variability survey in the field of the old,
super metal--rich open cluster NGC~6791. }
   {The data sample consists of more than 75,000 high--precision CCD time series measurements in  the $V$ band obtained mainly
 at the Canada--France--Hawaii Telescope, with additional data from S.~Pedro M\'artir and Loiano observatories,
 over a time span of ten nights. The field covers an area of $42 \times 28$ arcmin$^2$.}
   {We have discovered 260 new variables and re-determined periods and amplitudes of 70
known variable stars. By means of a photometric evaluation of the membership in NGC~6791,
and a preliminary membership based on the proper motions,
we give a full description of the variable content of the
cluster and surrounding field in the range 16\,$\lesssim V<$\,23.5.
Accurate periods can be given for the variables
with $P\lesssim$\,4.0~d, while for ones with longer periods the limited time--baseline hampered precise determinations.
We categorized the entire sample as follows: 6 pulsating, 3 irregular, 3 cataclysmic, 89 rotational variables and 61 eclipsing
systems; moreover, we detected 168 candidate variables for which we cannot give a variability class since
their periods are much longer than our time baseline.}
 {On the basis of photometric considerations, and of the positions of the stars
 with respect to the center of the cluster, we inferred that 11 new
 variable stars are likely members of the cluster, for 22 stars the membership
 is doubtful and 137 are likely non--members.
 We also detected an outburst of about 3 mag in the light curve of a very faint blue
 star belonging to the cluster and we suggest that this star could be a new U~Gem (dwarf nova) cataclysmic variable.}


    \maketitle
%

\begin{table*}
\caption{Previous variable star searches in NGC~6791.}
\label{tabprevious}
\centering

\begin{tabular}{l|r|l|l}
\hline \hline
 Authors                                                  & Nr. of variables & IDs & Notes \\
\hline
Kaluzny \& Rucinski (\cite{kaluzny93}) (KR93)             & 17  &  V1--V17                             & V15$\equiv$B7                                                                                             \\
Rucinski, Kaluzny \& Hilditch (\cite{rucinski96}) (RK96)  & 5   &  V18--V21 and B8		      &														   \\
Mochejska et al. (\cite{moch02}) (M02)                    & 47  &  V22--V67 and B4                     & B4 was previously catalogued by \\
 & & & Kaluzny \& Udalski (\cite{kaluzny92}) as \\
 & & & a blue star, but not as variable.  \\
Mochejska et al. (\cite{moch03}) (M03)                    & 7   &  V68--V74			      &														   \\
Kaluzny (\cite{kaluzny03}) (K03)                          & 4   &  V75--V78			      &														   \\
Bruntt et al. (\cite{bruntt03}) (B03)                     & 19  &  V79--V100                           & V85$\equiv$V76; V56$\equiv$V96; V77$\equiv$V88								   \\
Mochejska et al. (\cite{moch05}) (M05)                    & 14  &  V101--V114			      &														   \\
Hartman et al. (\cite{hartman05})                         & 10  &  V115--V124                          & Plus 7 suspected variables\\
\hline \hline
\end{tabular}
\end{table*}
\section{Introduction}

The photometric precision achieved by several ongoing transiting planet searches
 allows us to extend the census of variable stars down to very low amplitudes and faint
 magnitudes in selected sky regions.
Variable stars are an important source of astrophysical information:  from observations of them
 we are able to test several theories (e.g., evolutionary and pulsational models).
 Since all stars in an open cluster have essentially the same age, chemical composition
and distance, the study of variables which are cluster members can put more severe
 constraints on the physical parameters. Comparisons can also be made between
the variable stars of the cluster and those of the surrounding field.

In this paper, we present the study and the classification of 260 new variable stars
 that we found in the field of the open cluster NGC~6791, while for 70
 already known variables we compare our results with the previous ones.
 NGC~6791 ($\alpha$= 19$^h$ 20$^m$ 53$^s$; $\delta$= +37$^\circ$ 46\arcmin 18\arcsec),
 is a rich and well studied open cluster.
 It is thought to be the one of the oldest and probably the most metal-rich cluster known in our Galaxy.
 Its age is estimated to be about 8--9 Gyrs (Carraro et al., \cite{carraro06}; King et al., \cite{king05};
 Chaboyer, Green, \& Liebert, \cite{chaboyer99}; Stetson et al., \cite{stetson03};
 Kaluzny \& Rucinski, \cite{kaluzny95}); however,
the white dwarf cooling sequence indicates a different value, i.e., $\sim$2.4~Gyr (Bedin et al., \cite{bedin05}).
 The most recent estimates of its
 metallicity are [Fe/H]=+0.39 (Carraro et al., \cite{carraro06}),
 [Fe/H]=+0.47 (Gratton et al., \cite{gratton06}), and [Fe/H]=+0.45 (Anthony-Twarog et al., \cite{twarog06}).
In this work we adopt for NGC~6791 a distance modulus $(m-M)_{V}=13.35\pm0.20$ mag and a reddening
$E(B-V)$=0.09 mag (Carraro et al., \cite{carraro06}).
 The cluster is thus located at about 4.1 kpc from the Sun.

Because of its extreme characteristics,
 NGC~6791 has been the target of many surveys (see Table \ref{tabprevious}
 for a list of publications).
Taking into account the fact that in four cases the same stars have two identification numbers
 (V15$\equiv$B7, V56$\equiv$V96, V76$\equiv$V85 and V77$\equiv$V88) and
 counting also the stars B4 and B8, the total number of known variable stars in the
 field of NGC~6791 to date was 123 (plus 7 suspected variables, proposed by Hartman et al., \cite{hartman05}).

 In Sect.~\ref{observations} below we describe our observations,
 in Sect.~\ref{methods} we give details about the methods we employed
 in the search for variable stars, which are themselves presented in Sect.~\ref{variables}.
 In Sect.~\ref{inven} we describe the properties of the variable stars,
 focusing our attention on probable cluster members and some additional
 peculiar cases.
 The entire catalogue of variable stars is reported in an Appendix.

\section{Observations and data reduction}
\label{observations}
We surveyed NGC~6791 to detect the transits of extrasolar planets
 (Montalto et al., \cite{montalto}).
The campaign covered 10 consecutive nights (from July 4, 2002 to July 13, 2002)
and it was characterized by the continuous monitoring of the target on each clear
night. Therefore, in addition to the planetary transit search, we could get access
to the full variability content at $P\lesssim$\,4.0~d, both for the cluster and
the surrounding field.
Three telescopes were used:
\begin{enumerate}
\item The Canada--France--Hawaii Telescope (CFHT) in Hawaii equipped with the CFHT12k detector,
composed of 12 CCDs of 4128 $\times$ 2048 pixels and covering a field of
about 0.32 deg$^2$.
Owing to the large number of bad columns, data from chip 6 could not be used,
 so we could get data over a 0.29~deg$^2$ field;
\item The San San~Pedro M\'artir (SPM) 2.1--m telescope equipped  with the Thomson 2k detector
and covering a field of about $6~\times~6$~arcmin$^2$;
\item The Loiano 1.5--m telescope equipped with
BFOSC + the EEV 1300$\times$1348B detector and covering a field of $11.5~\times~11.5$~arcmin$^2$.
\end{enumerate}

\begin{figure*}[]
\centering
\includegraphics[width=1.4\columnwidth,height=1.2\columnwidth]{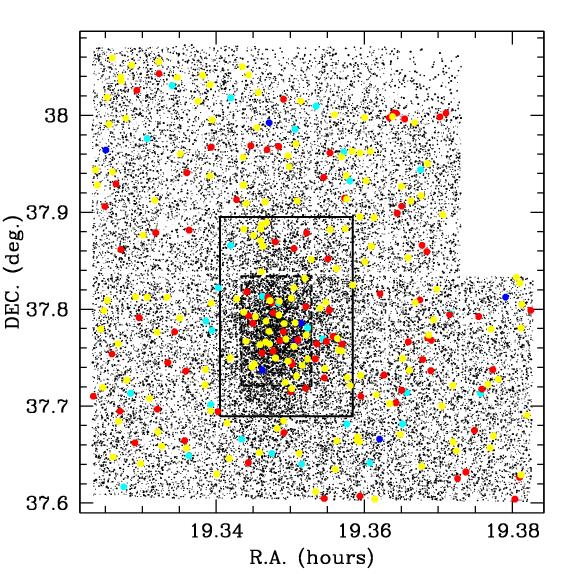}

\caption{\footnotesize Field of view (42 $\times$ 28 arcmin$^2$) of the CFHT image.
The chips are numbered in increasing order from left to right from chip~1 (top left) to chip~12 (bottom
right); stars of chip~6 are not plotted since we found it impossible to derive accurate
photometry.  Dashed and solid lines are the edges of the SPM and Loiano fields, respectively.
 Variable stars are also plotted: {\em blue dots} are pulsating variables, {\em purple dots}
are irregular and cataclysmic variables, {\em green dots} and {\em cyan dots} are
eclipsing systems (EA/EB--Type and EW--Type, respectively).
Finally, {\em red} and {\em yellow dots} are rotational and long--period variables, respectively.}
\label{fig_map}
\end{figure*}

Table~\ref{tab_oss} gives details about the length of the observing nights while
Figure \ref{fig_map} shows the field of the CFHT survey and the edges of the Loiano
and SPM surveys. The coordinates of the edges of our fields are also listed in Tab.~\ref{tab_oss}.
The field of the SPM observations is entirely included within chip 9 and
the field of the Loiano observations partially covers chips
2, 3, 4, 8, 9 and 10 of CFHT (see Fig.~\ref{fig_map}).
The luminosities of the new variables range from $V$=23.2~mag to $V\sim16$~mag
(near 1~mag above the turn--off); brighter stars are saturated.
The calibration of the CFHT, Loiano and SPM data have been performed
by using the Kaluzny \& Rucinski photometry (\cite{kaluzny95}).
More details on the data reduction procedure can be found in Montalto et al. (\cite{montalto}).

\begin{table}
\caption{The observation log for each night and the limits of the
field of view at the 3 different observatories.}
\label{tab_oss}
\centering
\begin{tabular}{c|cc|cc|cc}
\hline \hline
 &
 \multicolumn{2}{|c|}{Loiano}   &
 \multicolumn{2}{|c|}{SPM}    &
 \multicolumn{2}{|c}{CFHT} \\
\hline
 Night     & t$_{start}$      & $t_{end}$  & t$_{start}$      & $t_{end}$  & t$_{start}$      & $t_{end}$ \\ &
 \multicolumn{2}{|c|}{ (HJD--2452400)}   &
 \multicolumn{2}{|c|}{ (HJD--2452400)}   &
 \multicolumn{2}{|c} { (HJD--2452400)}   \\
\hline
1$^{st}$  &       &       &       &       & 59.82 & 59.96 \\
2$^{nd}$  &       &       &       &       & 60.83 & 61.02 \\
3$^{rd}$  & 62.48 & 62.63 & 61.68 & 61.95 & 61.94 & 62.07 \\
4$^{th}$  & 63.41 & 63.63 & 62.68 & 62.96 & 62.76 & 63.07 \\
5$^{th}$  & 64.38 & 64.64 & 63.68 & 63.96 & 63.88 & 64.10 \\
6$^{th}$  & 65.39 & 65.62 & 64.69 & 64.98 & 64.77 & 65.11 \\
7$^{th}$  &       &       & 65.70 & 65.82 &       &       \\
8$^{th}$  &       &       & 66.69 & 66.97 &       &       \\
9$^{th}$  &       &       & 67.67 & 67.98 & 67.77 & 68.10 \\
10$^{th}$ &       &       & 68.69 & 68.87 & 68.76 & 69.11 \\
\hline
$\alpha_{min}$ & \multicolumn{2}{|c|}{19$^h$ 20$^m$ 25\fs8}            & \multicolumn{2}{|c|}{19$^h$ 20$^m$ 36\fs3}             & \multicolumn{2}{|c}{19$^h$ 19$^m$ 23\fs7} \\
$\alpha_{max}$ & \multicolumn{2}{|c|}{19$^h$ 21$^m$ 30\fs4}            & \multicolumn{2}{|c|}{19$^h$ 21$^m$ 10\fs4}             & \multicolumn{2}{|c}{19$^h$ 22$^m$ 58\fs0}    \\
$\delta_{min}$ & \multicolumn{2}{|c|}{37$^\circ$ 41\arcmin 22\farcs6}  & \multicolumn{2}{|c|}{37$^\circ$ 43\arcmin 15\farcs4}   & \multicolumn{2}{|c}{37$^\circ$ 36\arcmin 6\farcs7}    \\
$\delta_{max}$ & \multicolumn{2}{|c|}{37$^\circ$ 53\arcmin 42\farcs7}  & \multicolumn{2}{|c|}{37$^\circ$ 50\arcmin  3\farcs1}   & \multicolumn{2}{|c}{38$^\circ$ 4\arcmin 21\farcs2}    \\
\hline
\hline
\end{tabular}
\end{table}

\section{The identification of variable stars}

\label{methods}
The intensive monitoring of NGC~6791 allowed us to obtain tens of thousands of photometric
time series for stars located in, close to, and far away from the cluster center.
We have analysed 73331, 6055, and 2152 light curves obtained from the CFHT, Loiano and SPM telescopes respectively.
The CFHT, Loiano and SPM time series are composed of about 250, 60 and 170 datapoints, respectively.
The observations, intended to detect photometric transits, were performed in the $V$ band only.

\subsection{The search for variable candidates}
To search for variable stars, we  calculated the best ``sinusoid plus constant" fit for
 all light curves (Vanicek, \cite{vani}; Ferraz-Mello, \cite{ferrazmello81}).
We evaluated the goodness of the fit by calculating
parameters related to the reduction of the initial variance obtained by introducing
the periodic term. These parameters are the reduction factor (Vanicek, \cite{vani}) and the
coefficient of spectral correlation  $S(\nu)$ (Ferraz-Mello, \cite{ferrazmello81}).

Owing to the huge number of light curves, we need one or more parameters to
discover the variability.
Toward this goal, we considered the parameter $r$ defined as
$r=\log_{10} S_{max}$, where $S_{max}$ is the maximum value of $S$ (i.e., the one corresponding to the frequency of the
 best-fit sinusoid in the Ferraz-Mello method). If a star does not show variability the
introduction of a sinusoid does not improve the fit and then $S_{max}$ is close
 to zero (no variance reduction) and $r\ll 0$; on the other hand,  a sine-shaped variability
strongly reduces the variance ($S$ close to 1) and hence  $r=0$. The purpose was to use the $r$ parameter
as a tracer of variability for short-period (i.e., intranight) variability.

To search for long--period variability, we introduced a second parameter, more sensitive to
the night--to--night variations.
We calculated the mean magnitude  $V_i$ and the standard deviation $\sigma_i$ on each night, and after that we calculated the parameter $s$ defined as:
\[
s=\log_{10} \frac{\Delta V }{\bar{{\sigma }}}
\]
where $\Delta V$ is the peak--to--peak difference
 and $\bar{\sigma}$ is the mean of the $\sigma_i$ over all nights.

\begin{figure}[]
   \centering
   \includegraphics[width=\columnwidth,height=\columnwidth]{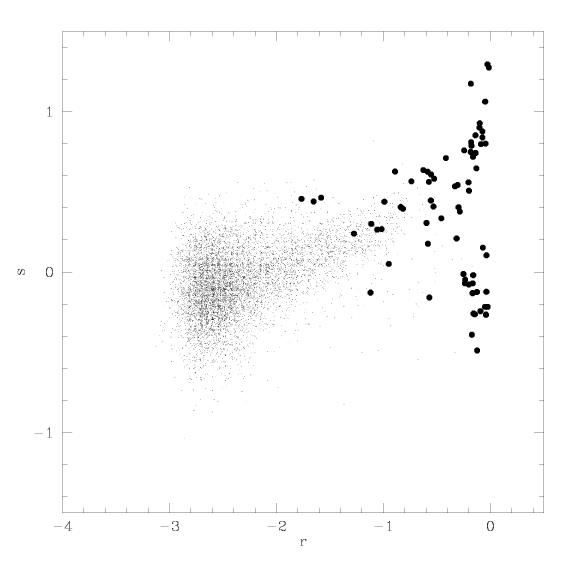}
   \caption{\footnotesize
 {\em Grey points}: $r$ parameter vs. $s$ parameter for constant
  light curves. {\em Black dots}: parameters derived from our light curves for the variable
 stars previously detected in our field.}
   \label{fig_rs}
   \end{figure}

To test the capability of the $r$ and $s$ parameters to detect variable stars,
we prepared a sample containing two types of light curves:
7722 artificial {\em constant} light curves (see Montalto et al., \cite{montalto}
for details) and 70 light curves of {\em already known variable stars} which
are included in  our CFHT field.
In Fig.~\ref{fig_rs} we plot $r$ vs. $s$ for the light curves
of constant stars (small  points) and of variable stars (large points).
The variable stars are substantially apart from the constant stars and
most have $r\gtrsim -1$. The variable stars with $r\lesssim -1$
and superposed on constant stars are mostly EA--type stars
 or irregular stars (e.g., cataclysmic variables).
Among variables (i.e., large dots in Fig.~\ref{fig_rs}), the stars with small $s$
have short periods ($P\leq 0.50$~d), while stars with large $s$ have long periods.
Therefore, we can conclude that the combination of the $r$ and $s$ parameters
is a good tracer of variability.

To detect the variable stars in our sample of $\sim$82,000 light curves we first
selected in an automatic way all the stars with $r\geq -2.0$, according to
the test described above. We thereby reduced the huge initial sample to $\sim$6,500 stars.
After the calculation of the amplitude spectrum of their
time series, we adopted as  a second selection criterion
a signal--to--noise ratio (S/N) greater than 4.0 around the highest peak in the
amplitude spectrum. This procedure allowed us to reduce our sample to $\sim$900
stars, i.e., 1.1\% of the whole initial sample.
Further checks have been made by examining the light curves of a random
sample of stars with $r<-2.0$, large $s$ and 3.5$<$S/N$<$4.0, but we did not
find any additional variables.

Our approach allowed us to detected hundreds of stars showing peaks in their
power spectra at $f=$1.00~cd$^{-1}$, at $f\leq$0.05~cd$^{-1}$, or
at $f$=0.6~cd$^{-1}$.
The first two spurious periodicities are common and can be ascribed to
small misalignments in the mean magnitudes from one night to the next
or recurrent drifts (caused by small color effects, for example) in the
intranight light curves. We suggest that the latter one is probably a
 photometric artefact occurring in some particular cases of blended stars,
or stars close to CCD edges, or bad pixels.
They have been considered as not reliable enough to infer a physical
light variability. In our opinion, only the combination of automatic
procedures and visual inspection allowed us to identify the three classes
($P$=1.00~d, $P$=1.6~d, $P\gg$10~d) of  spurious variables in the
huge number of $\sim$82,000 light curves. In
particular, we note that the
identification of the whole sample of eclipsing binaries has been confirmed by the
application of the box fitting technique (BLS, K\`ovacs et al. \cite{bls}),
used by Montalto et al. (\cite{montalto}) to detect planetary transits.

At the end of the variable star identification, we were left  with 330
cases to be characterized. Since we rejected about 2/3 of the
sample selected by means of the $r, s$ parameters, we are confident
we have not applied overly strict constraints in the candidate selection.

\subsection{The cross check with previous surveys of NGC~6791}

\label{variables}
When comparing our field of view with those
of other surveys, we found that 81 known variable stars are included.
The CFHT survey failed to detect 45 known variable stars:
 seventeen stars
(V22, V24, V26, V28, V30, V35, V36, V47, V50, V57, V61, V63, V64, V102, V103, V104, V105)
are outside the CFHT field of view;
4 stars (V71, V106, V113 and V120) lie between two chips;
23 stars (V1, V6, V13, V19, V33, V39, V45, V49, V54, V56$\equiv$V96, V65, V66, V67, V69,
V70, V72, V73, V74, V77$\equiv$V88, V78, V81, V97 and V112) are saturated; and the
V76 data are useless.

Among the 81 known variable stars that we have observed,
 not all of them display variability in our sample:
4 stars (V10, V18, V21, V32) are previously classified as long-period
detached eclipsing variables and we did not observed  eclipses.
We are not able to confirm the period of 15.24~days for V68 (M03),
likely because of our shorter time baseline and the small
amplitude of this variable (about 0.003~mag in $V$--band, M03).
Finally, we cannot confirm the variability
of six stars (V20, V79, V84, V98, V99, V116) and of the seven suspected variables
found by H05, since our data do not show any trace of variability.

Among the sample of the stars missing from the CFHT field,
 we identified 22 stars in the Loiano and SPM data sets
(V6, V13, V19, V20, V33, V45, V54, V56$\equiv$V96, V65, V66,
V67, V70, V71, V73, V74, V76$\equiv$V85, V77, V78, V81, V97, V106 and V113).
However, owing to the smaller signal--to--noise ratio (S/N),
the small number of datapoints and (in the case of the Loiano data)
the limited survey time, we could only confirm the variability
of stars V56$\equiv$V96, V66 and V76$\equiv$V85.

 Throughout this paper we use the existing names for the already known variables;
to identify the new ones discovered in our survey we used
the five--digit number assigned by the DAOPHOT package followed by the number
of the chip which the star belongs to. Accurate astrometry is provided to identify the stars
on the sky.  Moreover, all light curves of the variables 
will be available on CDS.

\section{The variable star content of NGC~6791 and its surrounding field}
\label{inven}

\begin{figure}[h]
\begin{center}
\includegraphics[width=\columnwidth,height=\columnwidth]{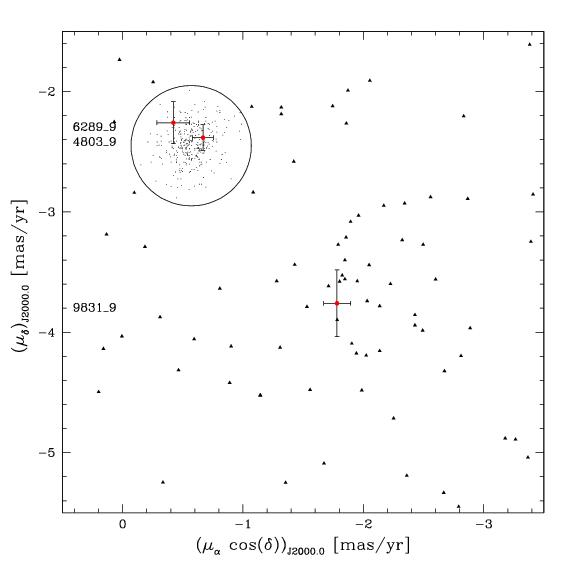}
\caption{\footnotesize
Proper-motion vector-point diagram for the inner region of NGC~6791 (from Bedin et al., \cite{bedin06}).
The circle (centered on the absolute proper motion of the cluster)
represents a safe limit corresponding to 0.5 mas/yr.
Triangles represent non--members, points with error bars are the new variables 06289\_9, 04803\_9 and 09831\_9}
\label{rolly}
\end{center}
\end{figure}

The CFHT measurements are quite precise, thus the light curves
are generally very well defined for $P<$4~d. On the other
hand, the periods and the shapes are  uncertain for $P>$4~d,
since  the observations only covered 2.5~cycles or less. 
We refer to Montalto et al. (\cite{montalto}) for a full description of the photometric
errors.  In order to evaluate the precision in the study of the variable stars,
we calculated the standard deviations of the Fourier least-squares fits (truncated
at the last significant term for the given star) for the 138 light curves having
very good phase coverage.
The precision was found to be  better than 0.010~mag in 73 cases (53\%), and better than 0.020~mag in
a total of 122 cases (88\%), as expected for stars ranging from $V\sim$16.0 to $V\sim$22.5.
The discussion based is mostly on the CFHT data, which are by far the most numerous, precise and
homogeneous; however, for some variables we have used data from Loiano and SPM in a very profitable way.
As an example, only the longitude spread of the three observatories allowed us
to derive the periods of the eclipsing binaries 00645\_10, V107, V12, V109
and of the rotational variable 03079\_9.

To proceed in the definition of the variable star content of NGC~6791 and its surrounding
field, we calculated
the power spectra of the data for all the 330 candidate variables
by using the least-squares iterative sine-wave search (Vanicek, \cite{vani}) and the Phase
Dispersion Minimization (Stellingwerf, \cite{stellingwerf}) methods.
Differences have been examined and resolved.  The separation
into different classes of variable stars has been made on the basis of the light
curve parameters (period, amplitude, Fourier coefficients) and standard photometric
values ($V$, $B-V$, $V-I$), when available.
The period estimates have been refined by means of a least--squares procedure
(MTRAP, Carpino \cite{carpino}) and appropriate error bars have also been
calculated. At the end of the process we get
six pulsating stars with $P<0.6$~d, three irregular variables, three cataclysmic variables (CVs), 31 detached or semi--detached
eclipsing binaries, 29 contact binaries, 90 rotational variables, 167 stars showing
clear night-to-night variability on
timescales too long for periods to be determined over our 9.2--d baseline.
We adopt preliminary membership probabilities based on proper motion measurements
kindly provided to us by K.~Cudworth (private communication) for 35 stars.
Moreover, for three new variable stars we adopted membership probabilities based on proper
motions performed by Bedin et al.~(\cite{bedin06}) (hereafter B06, see Figure~\ref{rolly}).

\begin{figure}[h]
\begin{center}
\includegraphics[width=0.75\columnwidth,height=0.75\columnwidth]{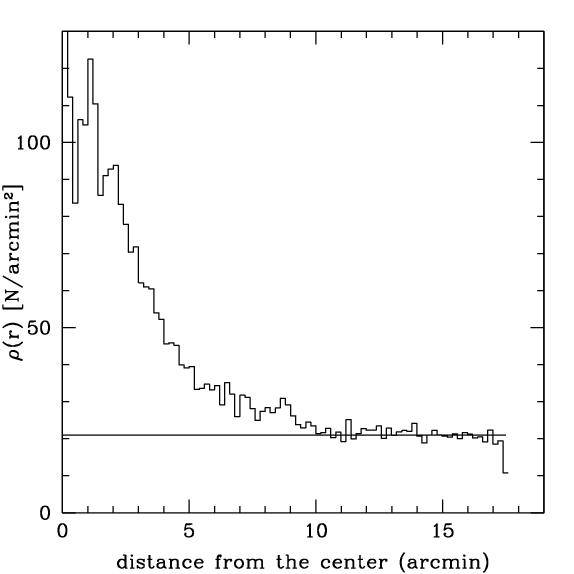}
\caption{\footnotesize Stellar density $\rho$ (number of stars per square arcminute)
 as a function of the distance from the center. The straight line represents the
 mean stellar density at distances greater than 10\arcmin.}
\label{distrad}
\end{center}
\end{figure}

For the other stars, we consider their position in color--magnitude diagrams (CMDs),
and their distance from the center of the cluster to infer whether they belong
to the cluster (for EW--Type stars we also utilize the $P$-$L$-$C$ relation of Rucinski (\cite{rucinski03}).
Toward this end, we plotted the radial distribution of all stars in Fig.~\ref{distrad}. We see that at a distance of $\sim$10\arcmin\,
 from the center of the cluster, the stellar density becomes near constant
(about 21~star/arcmin$^2$). Thus, we adopt the value of 10\arcmin\, as
 the external limit of the cluster and we consider ``likely non-members'' the variables
 located farther from the cluster center.

\subsection{Pulsating variables}
\begin{figure}[h]
\begin{center}
\includegraphics[width=0.30\columnwidth,height=0.30\columnwidth]{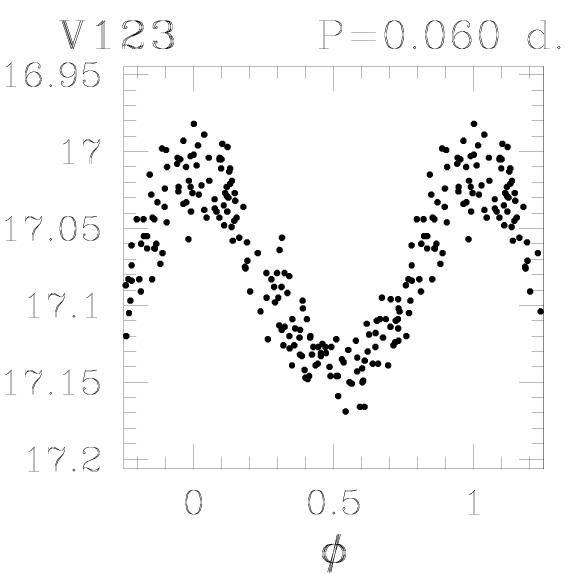}
\includegraphics[width=0.30\columnwidth,height=0.30\columnwidth]{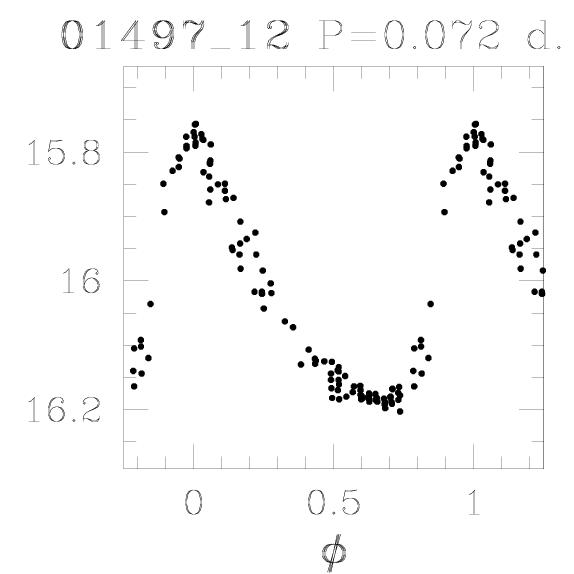}
\includegraphics[width=0.30\columnwidth,height=0.30\columnwidth]{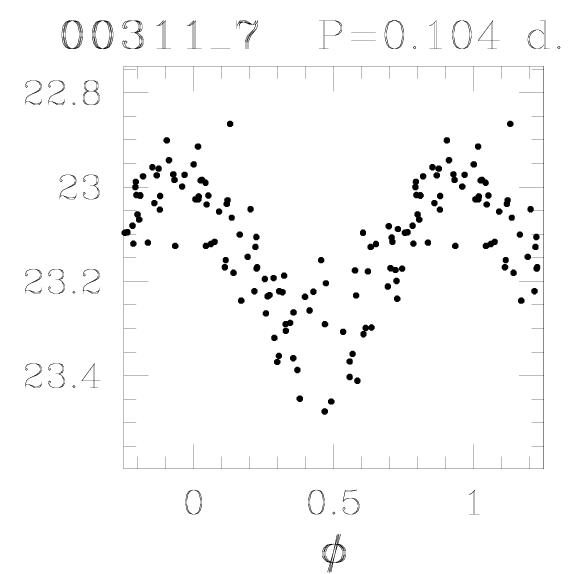}
\includegraphics[width=0.30\columnwidth,height=0.30\columnwidth]{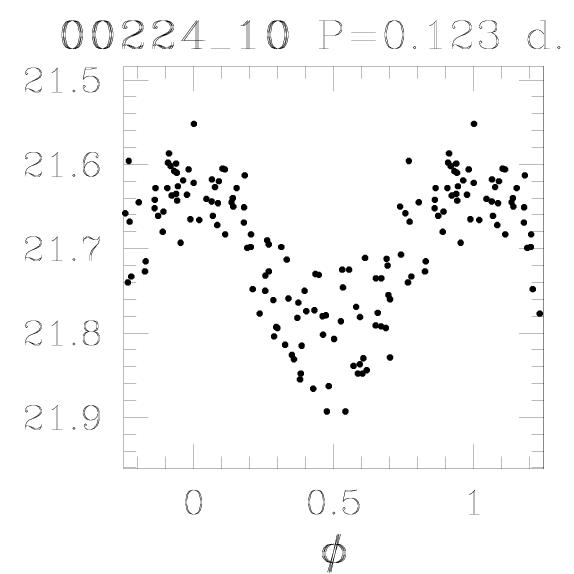}
\includegraphics[width=0.30\columnwidth,height=0.30\columnwidth]{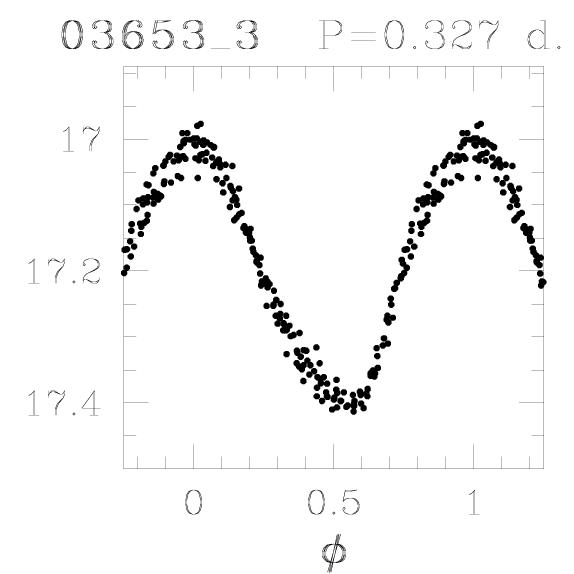}
\includegraphics[width=0.30\columnwidth,height=0.30\columnwidth]{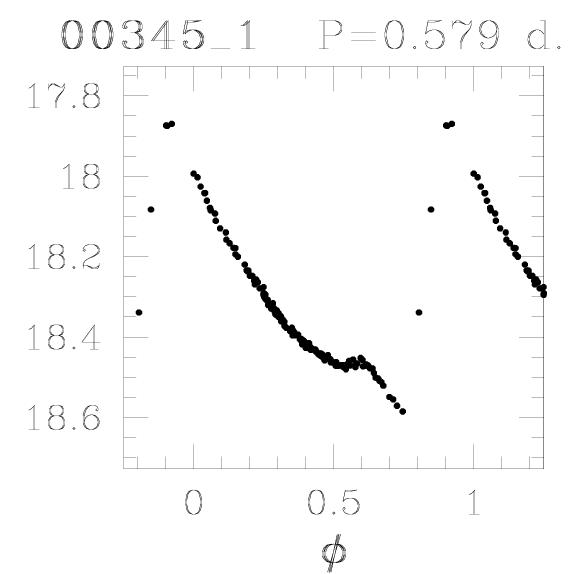}
\caption{\footnotesize Light curves of pulsating variables. V123 and 01497\_12 are probably HADS stars, 00311\_7 and 00224\_10 are SX Phe stars,
03653\_3 and 00345\_1  are RR Lyr stars.}
\label{lcpuls}
\end{center}
\end{figure}

The main characteristics of our variables are listed in Tab.~\ref{puls}
and their light curves are shown in Fig.~\ref{lcpuls}. The classification as
High--Amplitude Delta Sct (HADS), SX Phe, RRc or RRab stars
is based on the parameters of the Fourier decomposition
(Poretti, \cite{por01}). In all cases, the $\phi_{21}$ Fourier parameters
are on the progressions described by the different classes.
We note that our period for V123
is quite different from that given by H05 (0.107~d).
Error bars on the periods are in the range  1--6$\cdot10^{-5}$~d.

Both RR Lyr variables are too faint to belong to NGC~6791.
Since they have $V$=17.21 (03653\_3) and $V$=18.28 (00345\_1),
their distance moduli greatly exceed that of the cluster.

This is also true for the very faint and short-period stars
00311\_7 ($V$=23.17) and 00224\_10 ($V$=21.72);
therefore, it is more likely that they are Pop.~II stars.
On the other hand, using the $P-L$ relation given by McNamara (\cite{mcna}), we get
distance moduli of 14.50 and 13.78, respectively, for V123 and 01497\_12.
These distance moduli
and the distance from the cluster center (12\arcmin\, and 22\arcmin, respectively)
suggest that they do not belong to the cluster, though they are not very far from it.
Therefore, they are probably Pop.~I
stars and hence High Amplitude $\delta$~Scuti stars.

Moreover, there are several variables whose light curves
are very similar to those of Cepheid variables;
the Fourier decomposition of some light curves  (in particular the
large amplitude ones, i.e., 00913\_5, 01659\_8, V46 and 01431\_10,
but also 01606\_11, 02285\_10, 00122\_4 and  03056\_3) yields
parameters typical for Cepheid light curves. However, most of these variables are quite
faint and the Period--Luminosity relation for Cepheids (Tammann et al., \cite{tsr03})
yields distances in the range 39-171~Kpc.
It is difficult to say whether these stars are nearby rotational
variables (see below) in the Milky Way or very distant pulsating variables.  For our
present purposes, these stars have been included among the rotational variables listed
in the Appendix.

The puzzling nature of all these apparently distant stars (i.e., the Cepheid-like ones, the
RR Lyr and the faint SX Phe variables discussed above)
deserves further investigation by means of spectroscopic and/or kinematic data.

\subsection{Irregular variables}

Table~\ref{puls} also lists three irregular variables: these stars lie on the
middle Main Sequence and are all located less than 3\arcmin\, from the cluster center; thus we suggest that they belong to the cluster.
V92 and V83 were previously defined as ``periodic variables" by B03.
Indeed, we noticed fast variability in our light curves
(Fig.~\ref{irr}), but,
more noticeably, the mean magnitude is also changing from night to night.
The long periods given by M03 are not able to explain either the short- or the
longer-timescale variability; actually, we could not detect any periodic term.
We also detected no trace of periodicity in V93 (Fig.~\ref{irr}); we suspect
that the periods given by M05 and B03 are spurious, since they are close to 1.0~d (0.99 and 0.94~d,
respectively) and they could be produced by the irregular fluctuations.

We can conjecture that these variables are eruptive variables observed
in a quiescent phase, in which rapid and/or slow
changes with smaller amplitude can be observed; they resemble the case of V15
(see Sect.~\ref{cata}). We have no reliable indications about the membership probabilities.

\subsection{Cataclysmic variables}\label{cata}
\begin{figure}[h]
\begin{center}
\includegraphics[width=0.8\columnwidth,height=0.99\columnwidth]{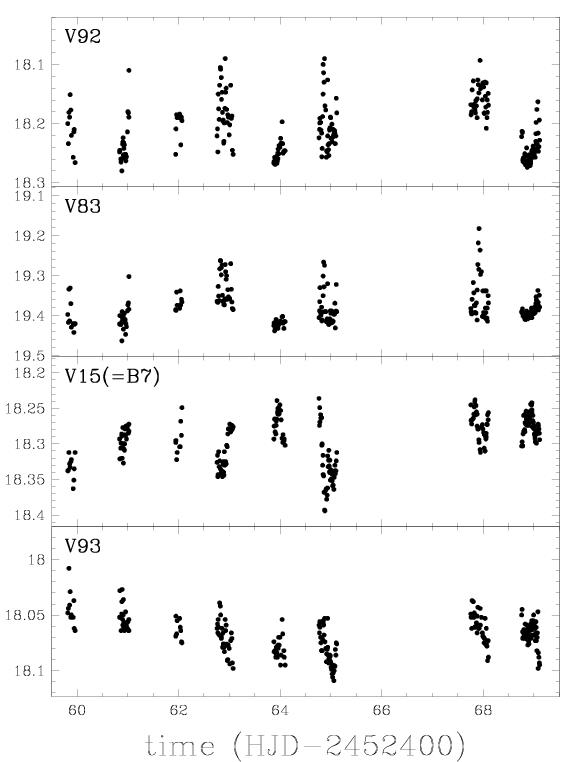}
\caption{\footnotesize Variable stars showing irregular fluctuations.}
\label{irr}
\end{center}
\end{figure}

As regards V15: M03 and M05 detected variability over the range
of 3 mag and observed outbursts of about 0.5-1.0 mag; from our side,
we could see a 0.15--mag variability in our light curves (Fig.~\ref{irr}),
corresponding to the quiescent phase. V15 is very probably a NGC~6971 member,
since the Cudworth proper-motion membership probability is very high (98\%).

\begin{figure}[h]
\begin{center}

\includegraphics[width=0.75\columnwidth,height=0.75\columnwidth]{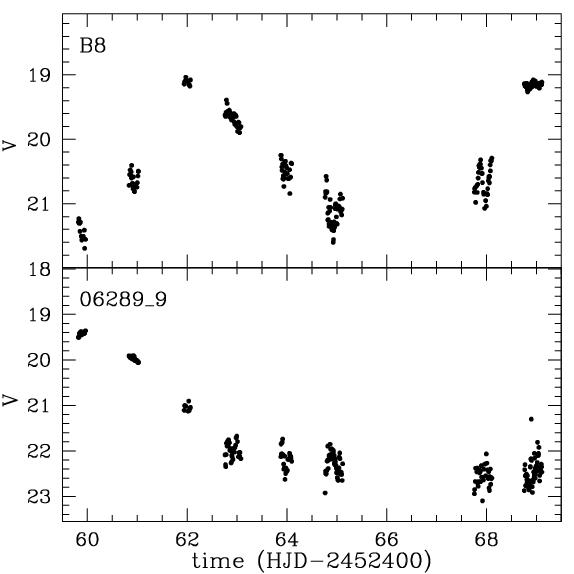}
\includegraphics[width=0.75\columnwidth,height=0.75\columnwidth]{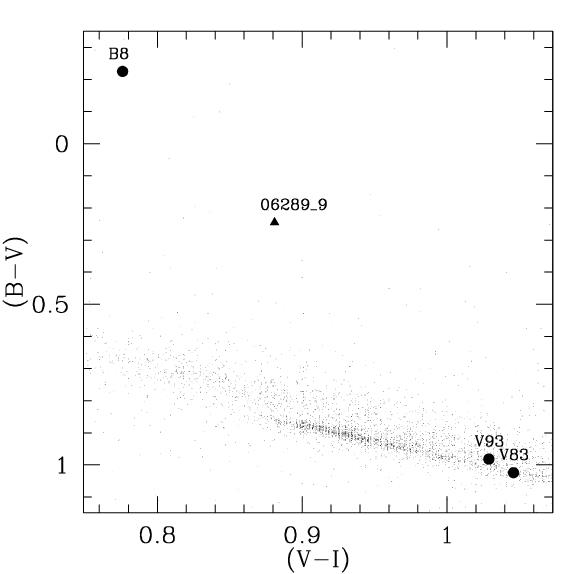}
\caption{\footnotesize {\em Top}: Light curves of B8 (CV star) and 06289\_9 (candidate CV).
                       {\em Bottom}: positions in the two--colour diagram for two irregular
                        stars (V83 and V93), for the cataclysmic variable B8 when in a low state,
                        and for the new variable candidate 06289\_9.}

\label{cv}
\end{center}
\end{figure}

Both the position of the faint blue star 06289\_9 in the two-colour diagram
and the shape of its
light curve (Figure~\ref{cv}) strongly suggest that  this star could be a new cataclysmic variable
(U~Gem-type, dwarf nova). Moreover, we know that this object is a cluster member
(see Figure~\ref{rolly}).
The star shows an outburst of about 3~mag and, though we did not observe the entire brightening,
 we would highlight that the magnitude was still increasing on the first night;
 thus we are able to say that the maximum brightness was reached immediately after.

We can estimate the  orbital period, $P_{orb}$, and the
recurrence time, $T_n$, from the decay
 time, $\tau_d$=$\Delta t /\Delta m$ [days~mag$^{-1}$] and the amplitude,
 $\Delta m$ (Warner, \cite{warner}, equations 3.5, 3.1, respectively).
 Assuming for $\Delta t$ and $\Delta m$ the values 3.33$\pm$0.50 d and 2.87$\pm$0.31 mag
 respectively, we find
 $P_{orb}$=2.54$\pm$1.41~h and  $T_n$=13.9$\pm$10.6~d.
However, our light curve (Fig.~\ref{cv}) seems to rule out
$T_n$ values shorter than 8~d.

The variable B8  shows a large-amplitude light curve (Fig.~\ref{cv}) over
a quite short  7~d time span. The cataclysmic nature of B8
has been confirmed spectroscopically by Kaluzny et al. (\cite{kaluzny97}) who
also notes that B8 exhibits red $V-I$ colour while in a low state.

Following the same procedure used for 06289\_9 and  assuming
$\tau_d$=1.3$\pm$0.3~d~mag$^{-1}$ for B8,
we find $P_{orb}$=2.97$\pm$1.63~h and $T_n$=11.4$\pm$8.5~d.
The $T_n$ value is compatible with the 7~d periodicity (Fig.~\ref{cv}).
A membership probability is not available for B8.
However, using the equations 3.3 and 3.4 after Warner (\cite{warner}), we obtain
$M_{Vmin}$=8.06$\pm$0.68~mag  and $M_{Vmax}$=4.97$\pm$0.42~mag.
In turn, these values give two estimates for the distance modulus of B8, i.e.,
13.82~$\pm$~0.68~mag and 14.20~$\pm$~0.42 mag. We note that the first is in agreement with
the distance modulus of the cluster.
Kaluzny et al. (\cite{kaluzny97}) assumed that B8 belongs to the cluster,
 finding $M_{Vmax}$=5.2 mag and $M_{Vmin}$=7.6, i.e., values very similar to ours.
B8 is located at 4\arcmin\,from the center, and we can only conclude that the
membership of this star is very probable.

\begin{table*}
\begin{flushleft}
\caption{Pulsating, irregular and cataclysmic variables. $V$ is the minimum brightness for
CVs and irregular, the mean brightness for pulsating variables. $T_0$ is the
time of maximum brightness for pulsating stars.
Hereafter, the labels ``k" and ``s" indicate that the $B-V$ color
index is taken from Kaluzny \& Rucinski (\cite{kaluzny95} or Stetson et al. (\cite{stetson03}), respectively.}

\begin{tabular}{ll ll ccc llll l}
\hline \hline
\noalign{\smallskip}
\multicolumn{1}{l}{Star} &
\multicolumn{1}{l}{Type}&
\multicolumn{1}{l}{$\alpha_{2000}$}&
\multicolumn{1}{l}{$\delta_{2000}$}&
\multicolumn{1}{c}{$V$} &
\multicolumn{1}{c}{$<B-V>$} &
\multicolumn{1}{c}{$<V-I>$} &
\multicolumn{1}{l}{Ref.} &
\multicolumn{1}{l}{$T_0$ }&
\multicolumn{1}{l}{Period}&
\multicolumn{1}{l}{Ampl.}  \\
 & & & & [mag] & [mag] & [mag] & & [HJD--2452400] & [d] & [mag]  \\
\noalign{\smallskip}
\hline
\noalign{\smallskip}

\multicolumn{10}{c}{Pulsating variables}\\
\noalign{\smallskip}

      V123 &   HADS &  19.362064 &  37.666034 &   17.08 &    0.45  &         &    k & 59.559 &    0.06026 &  0.14   \\
 01497\_12 &   HADS &  19.379083 &  37.812419 &   16.06 &          &         &      & 59.528 &    0.07227 &  0.40   \\
  00311\_7 &  SXPhe &  19.324628 &  37.716768 &   23.17 &          &         &      & 59.605 &    0.10443 &  0.10   \\
 00224\_10 &  SXPhe &  19.353639 &  37.710163 &   21.72 &    0.71  &    1.06 &    s & 59.801 &    0.12261 &  0.20   \\
  03653\_3 &    RRc &  19.347147 &  37.992413 &   17.21 &    0.57  &    0.58 &    k & 59.937 &    0.32654 &  0.39   \\
  00345\_1 &   RRab &  19.325082 &  37.964170 &   18.28 &          &         &      & 60.151 &    0.57866 &  0.72   \\
\noalign{\smallskip}
\noalign{\smallskip}
\noalign{\smallskip}
\multicolumn{10}{c}{Irregular  variables}\\
\noalign{\smallskip}

       V92 &    IRR &  19.350754 &  37.766876 &   18.10 &    0.91  &         &    k &            &            & 0.10   \\
       V83 &    IRR &  19.346220 &  37.737232 &   19.10 &    1.02  &    1.05 &    k &            &            & 0.07   \\
       V93 &    IRR &  19.351452 &  37.785687 &   18.12 &    0.98  &    1.03 &    s &            &            & 0.04   \\

\noalign{\smallskip}
\noalign{\smallskip}
\noalign{\smallskip}
\multicolumn{10}{c}{Cataclysmic variables}\\
\noalign{\smallskip}

  V15(=B7) &     CV &  19.352057 &  37.799019 &   18.26 &    0.20  &         &    k &            &            & 0.06   \\
        B8 &     CV &  19.343262 &  37.747833 &   20.64 &  --0.23  &    0.78 &    k &            &            & 2.27   \\
  06289\_9 & CV (?) &  19.348976 &  37.770355 &   22.80 &    0.25  &    0.88 &    s &            &            & 3.10   \\

\noalign{\smallskip}
\hline \hline
\label{puls}

\end{tabular}
\end{flushleft}
\end{table*}

\subsection{Contact binaries}

\begin{figure}[h]
\begin{center}
\includegraphics[width=0.30\columnwidth,height=0.30\columnwidth]{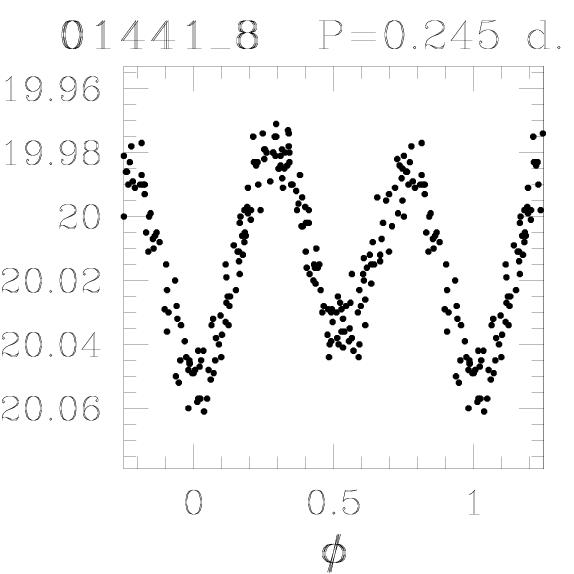}
\includegraphics[width=0.30\columnwidth,height=0.30\columnwidth]{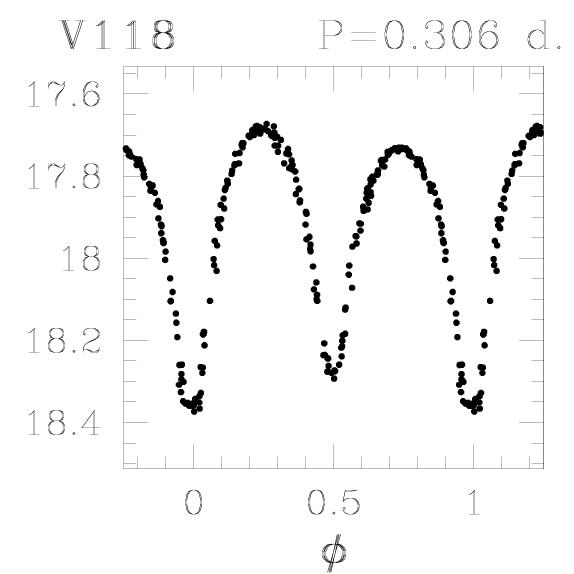}
\includegraphics[width=0.30\columnwidth,height=0.30\columnwidth]{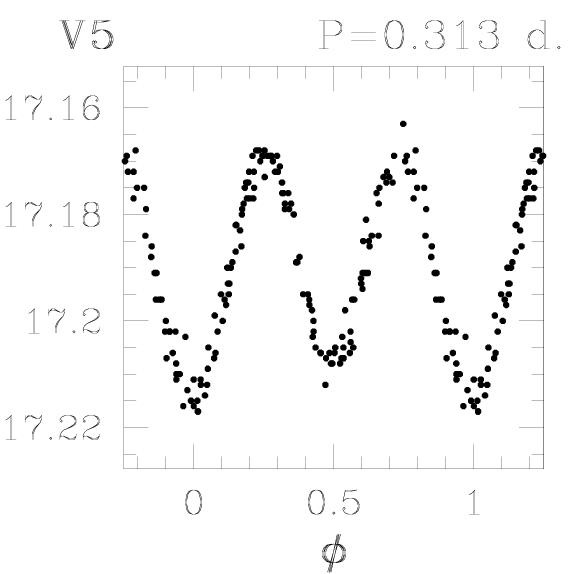}
\includegraphics[width=0.30\columnwidth,height=0.30\columnwidth]{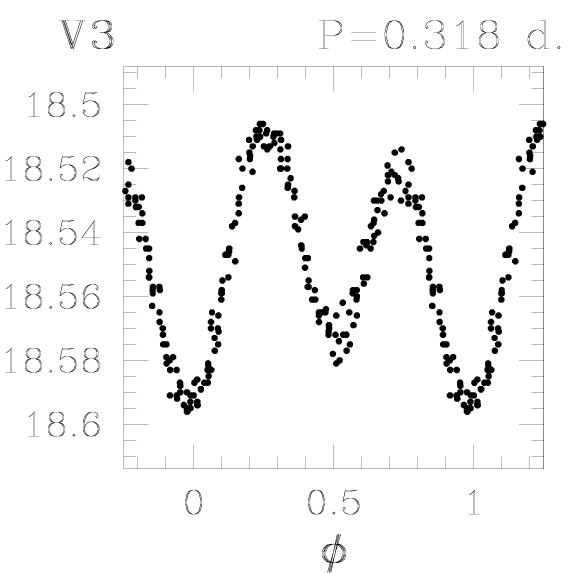}
\includegraphics[width=0.30\columnwidth,height=0.30\columnwidth]{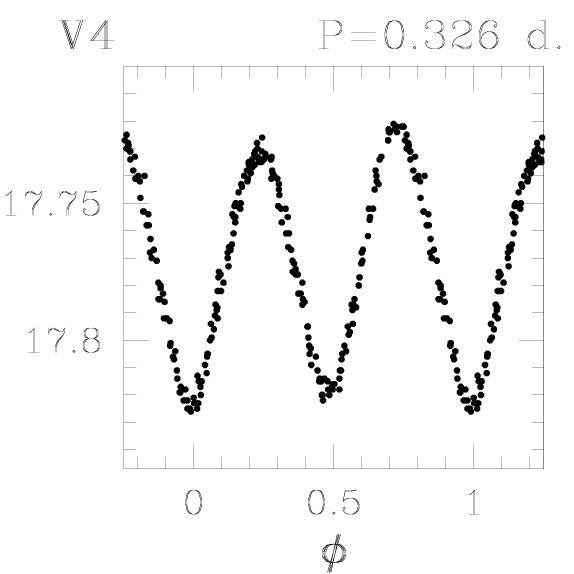}
\includegraphics[width=0.30\columnwidth,height=0.30\columnwidth]{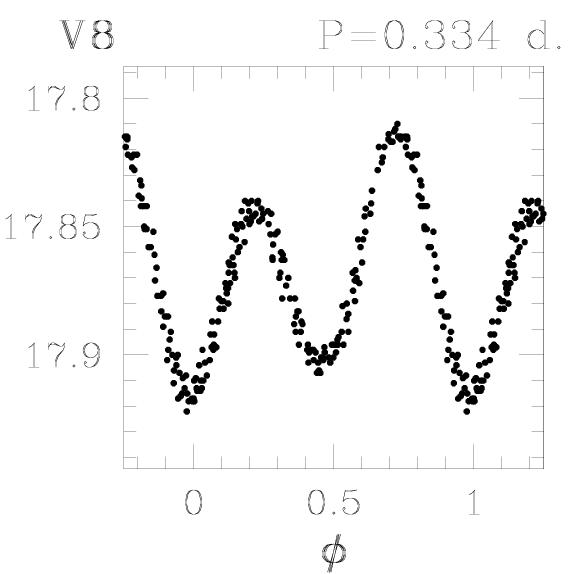}
\includegraphics[width=0.30\columnwidth,height=0.30\columnwidth]{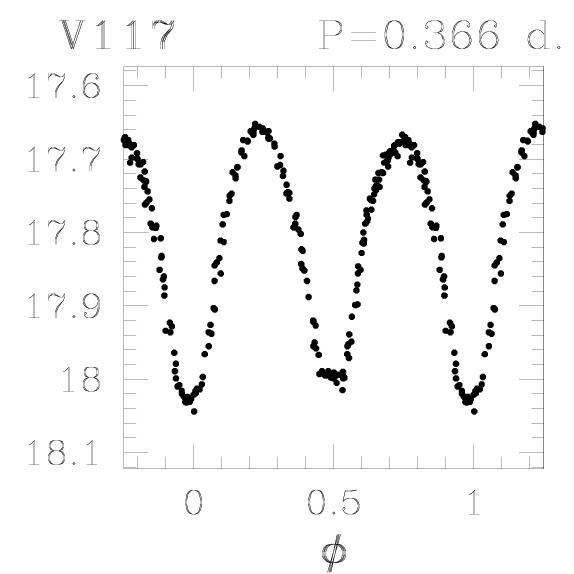}
\includegraphics[width=0.30\columnwidth,height=0.30\columnwidth]{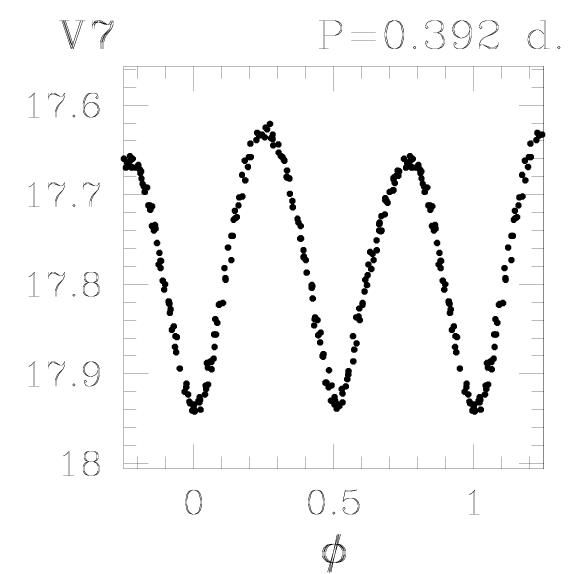}
\includegraphics[width=0.30\columnwidth,height=0.30\columnwidth]{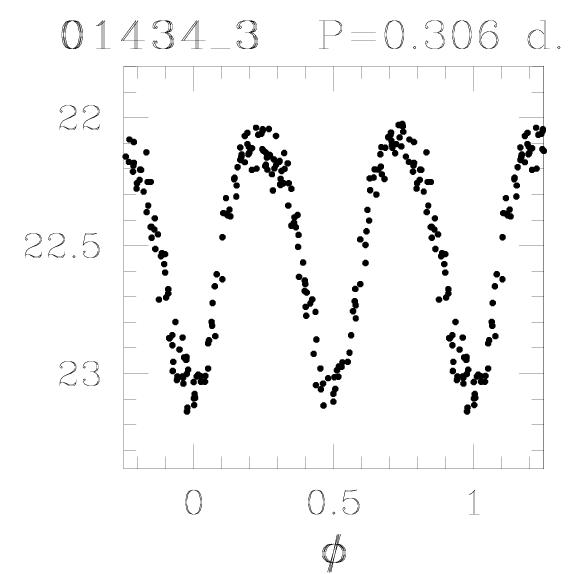}
\caption{\footnotesize The light curves of the contact binaries that are
likely members of NGC~6791. The case of 01434\_3 (amplitude much larger than
0.75~mag) is also shown in the last panel.}
\label{lcewbel}
\end{center}
\end{figure}

   \begin{figure}[h]
   \centering
   \includegraphics[width=0.99\columnwidth,height=0.99\columnwidth]{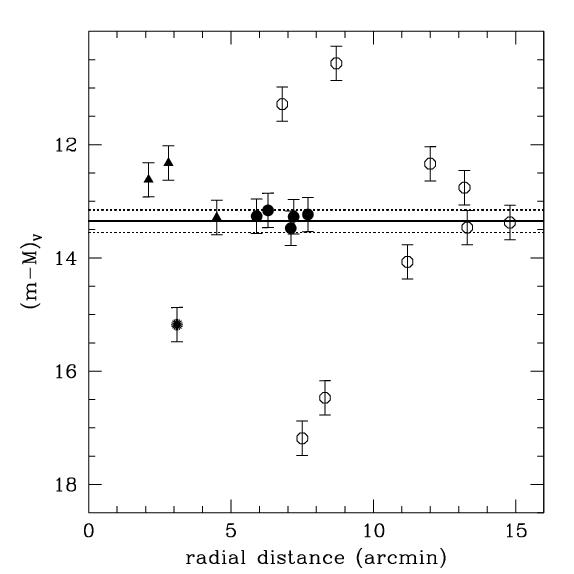}
   \caption{\footnotesize
The $m-M$ values calculated for contact binaries
by means of the  Rucinski (\cite{rucinski03}) $P$-$L$-$C$ relation plotted against the distance from
the cluster center. Triangles indicate the stars whose membership has been proposed by
M03, the filled circles the stars whose membership has been proposed by us, and the open circles the stars
that we suggest do not belong to NGC~6791.
The starred point indicates the star 09891\_9  which does not belong to the cluster
on the basis of the proper motion (B06).
The $(m-M)_V$ value for the cluster (solid line) with an error bar of $\pm$0.20 mag
(dashed lines) is also shown.
}
\label{membri}
\end{figure}

The simplest cases of eclipsing systems are the contact binaries
(also named W UMa systems); they show
short periods  and continuous variability
 and therefore can be easily recognized and classified. We
detected 29 of these variable stars; they have
$P<$0.40~days and very well defined light curves. The complete list and the light curves are reported in the Appendix.
Tab.~\ref{ewbel} lists the stars likely belonging to NGC~6791 (see above);
their light curves are shown in Fig.~\ref{lcewbel}.
The very short periods and the secondary minima occurring at $\phi \approx 0.5$
 indicate binaries with circular orbits, as is also the case for stars with small amplitudes (in 7 cases
we have amplitudes less than 0.20~mag: V3, V4, V5, V8, V23, V40 and 01441\_8).
The average error bar on the period estimates is of the order of 4--5$\cdot10^{-5}$~d.

However, we note that the stellar surfaces
are not homogeneous since the maxima are often at different heights. Therefore, binarity and
activity are probably combined here.
In particular, the shape of the light curves of V4 (comparing RK96, M02 and our data)
and V7 (comparing K93 and our data) have changed a lot;
we suppose that stellar spots strongly modify the light curves.
Proximity effects are also responsible for
the large amplitudes observed for  01434\_3 (Fig.~\ref{lcewbel}, last panel) and 00766\_5.
We also found different periods for V118 (0.306321~d) and V124 (0.320143~d) compared to H05.

As for membership, the probabilities provided by Cudworth are
78\%, 98\% and 98\% for V3, V4 and V5, respectively. Indeed, they are very close
to the cluster center (4\farcm5, 2\farcm1 and 2\farcm8, respectively).

We suggest that
01441\_8, V118, V8, V117 and V7 are also contact binaries belonging to NGC~6791.
To further verify this hint,  we calculated their distance by using the $P$-$L$-$C$ relation given by Rucinski (\cite{rucinski03}); they
turn out to have distance moduli
(13.28, 13.28, 13.48, 13.28 and 13.18, respectively) very similar to that of the cluster
(13.35).

Moreover, these stars  are located at similar angular distances from the cluster center 
(6\farcm2, 7\farcm2, 7\farcm1, 7\farcm2 and 6\farcm3, respectively).
Figure~\ref{membri}
shows how the distance modulus of the cluster is in better agreement
with those of the stars we proposed as cluster members than with those of the previously known members.
Their positions in the CMDs (Fig.~\ref{cmde}, filled circles) are similar
to those of stars in the sample with $V<$7.5 whose parallaxes have been determinated by HIPPARCOS
(Rucinski, \cite{rucinski03}). 
We also note that most of the cluster members are near the turnoff point.

\begin{table*}
\begin{flushleft}
\caption{Coordinates and light curve parameters of the contact binaries (W UMa systems, EW)
 belonging to NGC~6791. $V$ is the brightness at maximum
and  $T_0$ is the time of the primary minimum.}
\begin{tabular}{l ll ccc l ll ll}
\hline \hline
\noalign{\smallskip}

\multicolumn{1}{l}{Star} &
\multicolumn{1}{l}{$\alpha_{2000}$}&
\multicolumn{1}{l}{$\delta_{2000}$}&
\multicolumn{1}{c}{$V$} &
\multicolumn{1}{c}{$<B-V>$} &
\multicolumn{1}{c}{$<V-I>$} &
\multicolumn{1}{l}{Ref.} &
\multicolumn{1}{l}{$T_0$}&
\multicolumn{1}{l}{Period}&
\multicolumn{1}{l}{Ampl.} &
\multicolumn{1}{l}{Notes}  \\
 & & & [mag] &[mag] & [mag] & &
\multicolumn{1}{l}{[HJD--2452400]}&
\multicolumn{1}{l}{[d]}&
\multicolumn{1}{l}{[mag]} &  \\

\noalign{\smallskip}
\hline
\noalign{\smallskip}

  01441\_8  &   19.339422 &  37.778118 &   19.98 &    1.38  &         &    k & 63.073  &    0.24544 &  0.07 &         likely memb.  \\
      V118  &   19.347500 &  37.651222 &   17.68 &    0.75  &    1.01 &    s & 59.912  &    0.30623 &  0.70 &         likely memb.  \\
        V5  &   19.346258 &  37.813354 &   17.19 &    0.90  &    0.95 &    k & 60.221  &    0.31274 &  0.05 &        member (98\%)  \\
        V3  &   19.354380 &  37.769349 &   18.51 &    1.05  &    1.06 &    k & 59.798  &    0.31764 &  0.09 &        member (78\%)  \\
        V4  &   19.348396 &  37.806652 &   17.72 &    1.01  &         &    k & 59.591  &    0.32568 &  0.10 &        member (98\%)  \\
        V8  &   19.341938 &  37.865810 &   17.81 &    0.79  &    0.88 &    k & 59.896  &    0.33406 &  0.10 &         likely memb.  \\
      V117  &   19.343433 &  37.665848 &   17.66 &    0.87  &    0.90 &    k & 59.987  &    0.36644 &  0.38 &         likely memb.  \\
        V7  &   19.340271 &  37.821892 &   17.63 &    0.93  &    0.86 &    k & 59.820  &    0.39174 &  0.31 &         likely memb.  \\

\noalign{\smallskip}
\hline \hline
\label{ewbel}
\end{tabular}
\end{flushleft}
\end{table*}

\subsection{Eclipsing variables}

\begin{figure}[h]
\begin{center}
\includegraphics[width=0.30\columnwidth,height=0.30\columnwidth]{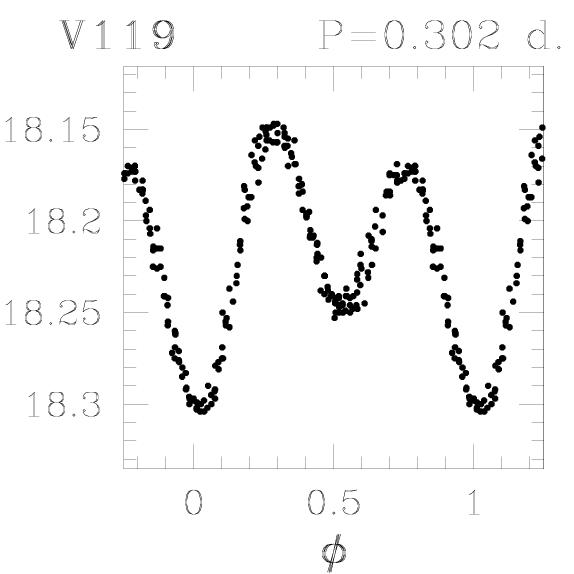}
\includegraphics[width=0.30\columnwidth,height=0.30\columnwidth]{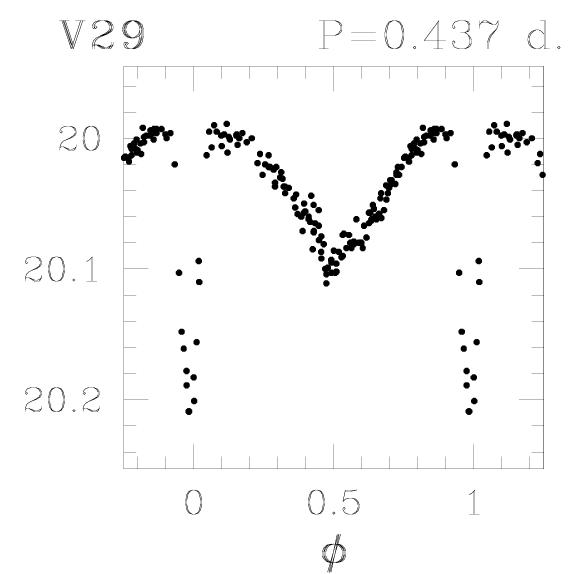}
\includegraphics[width=0.30\columnwidth,height=0.30\columnwidth]{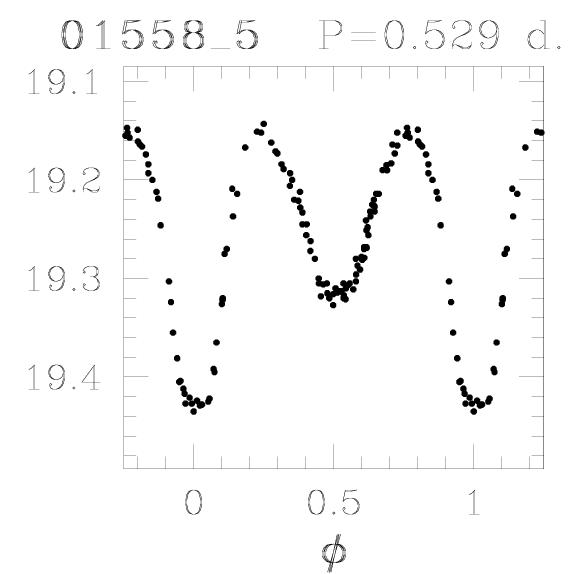}
\includegraphics[width=0.30\columnwidth,height=0.30\columnwidth]{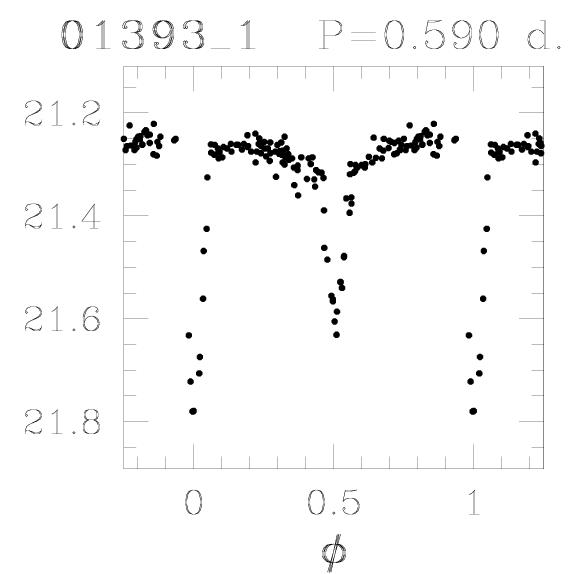}
\includegraphics[width=0.30\columnwidth,height=0.30\columnwidth]{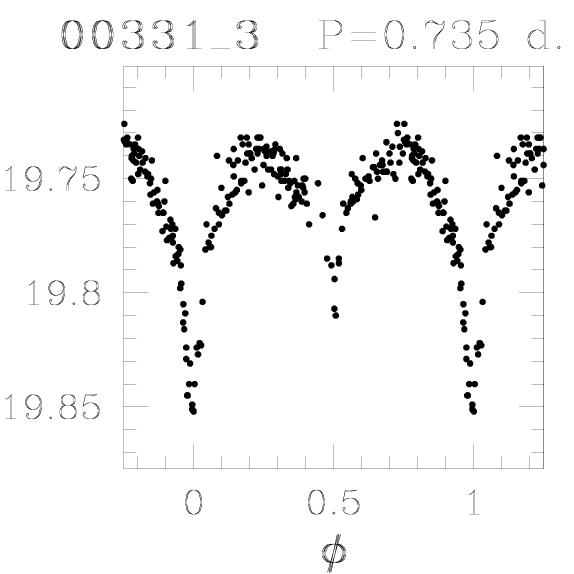}
\includegraphics[width=0.30\columnwidth,height=0.30\columnwidth]{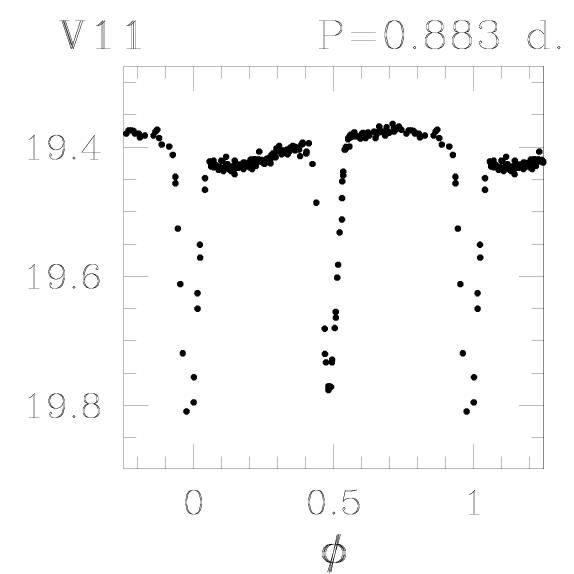}
\includegraphics[width=0.30\columnwidth,height=0.30\columnwidth]{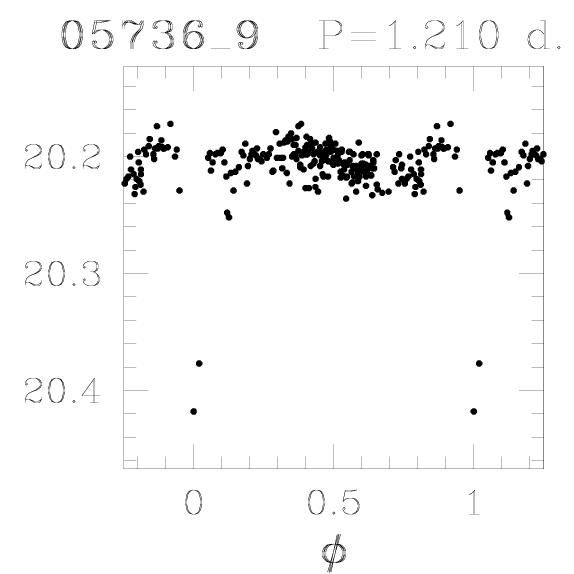}
\includegraphics[width=0.30\columnwidth,height=0.30\columnwidth]{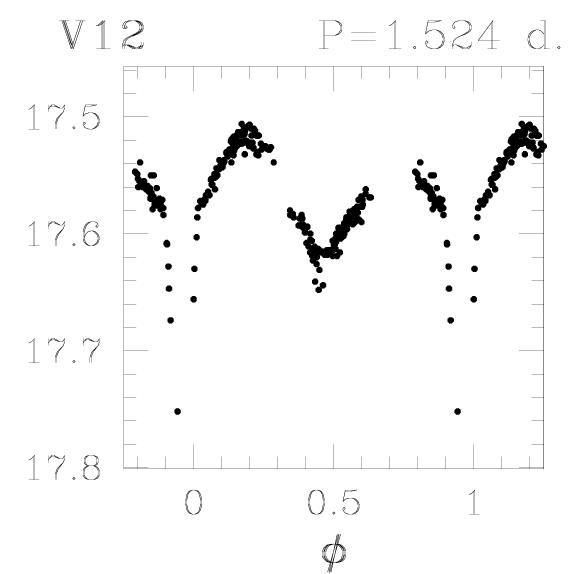}
\includegraphics[width=0.30\columnwidth,height=0.30\columnwidth]{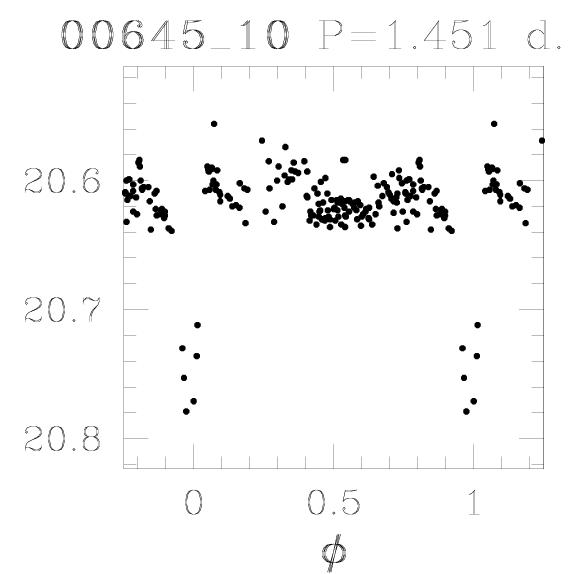}
\includegraphics[width=0.30\columnwidth,height=0.30\columnwidth]{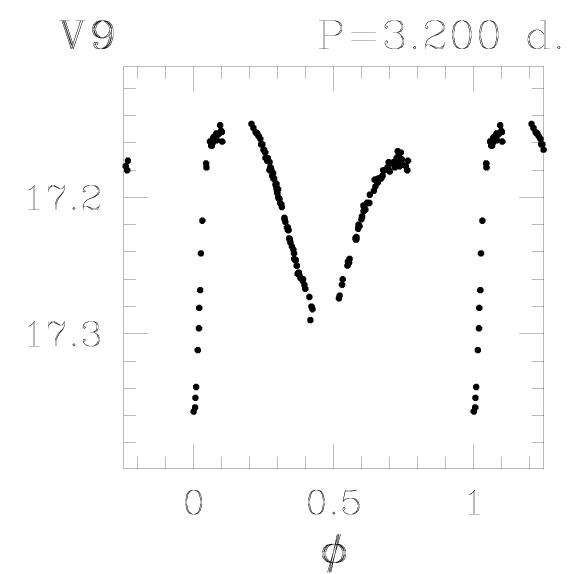}
\includegraphics[width=0.30\columnwidth,height=0.30\columnwidth]{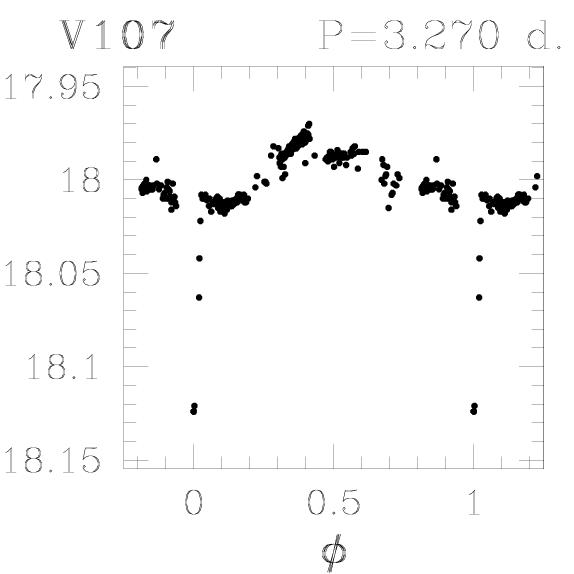}
\includegraphics[width=0.30\columnwidth,height=0.30\columnwidth]{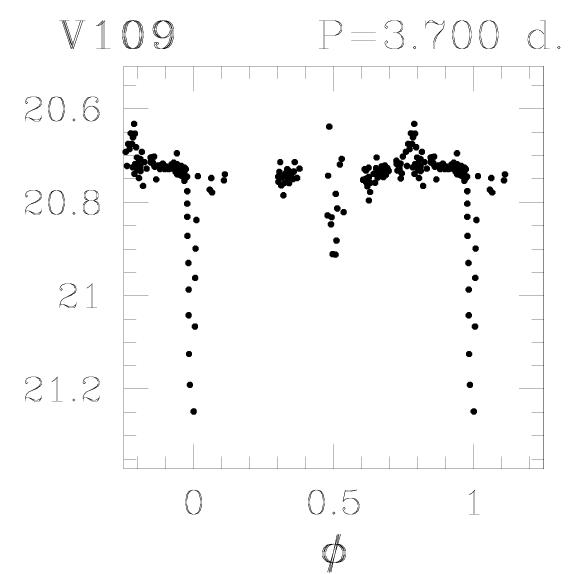}
\caption{\footnotesize The light curves of short--period detached or semi--detached
eclipsing binaries.}
\label{lcea}
\end{center}
\end{figure}

In the cases of detached or semi--detached eclipsing binaries the classification
and membership tasks are different from the case of
contact binaries.
Tab.~\ref{tabea} lists the systems for which we could determine periods; their light curves
are shown in Fig.~\ref{lcea}.
We still have short-period cases where we can reconstruct the complete light curve, as
for the classical examples of $\beta$ Lyr variables (V29, 01558\_5 and 00331\_3).
V9 is a  more complicated $\beta$ Lyr system in which  spots produce maxima with different
heights. Indeed, it has been classified as an RS~CVn variable by
M05 and B03; they also observed a ``shift of the modulation wave" from 1995 to 2002.

We note that our period for V119 is  quite different from that given by H05
(0.1133~days); the new period makes this star an intermediate case between semi--detached and contact systems.
Error bars on the periods in Tab.~\ref{tabea} are $\sim10^{-4}$~d for $P<$1.0~d, $\sim10^{-3}$~d for $1.0<P<$2.0~d
and a bit larger for $P>$3.0~d.

Some variables show very sharp eclipses and out-of-eclipse variability due to different levels of
stellar activity
(05736\_9, 00645\_10, V109, 01393\_1, V11 and V107; for the period of the latter star we prefer the longer of the two values given by M05).

In many cases we observed  one eclipse only and we cannot give any value for
the period, unless it has been given in the  previous studies, as for the
cluster member  V80
(86\% on the basis of the Cudworth membership probability).
We also note that the amplitude we observed in V80 is much larger than that
reported by B03.

To establish the membership of these eclipsing systems is not an easy task, since binary effects
should be taken into account when considering colors and  magnitudes. However,
on the basis of the distance from the cluster center and their position
in the CMDs, we can argue that V60, 02461\_8 (both single-event eclipsing
binaries), 05736\_9, V29 and 00645\_10
are very probable members. This hint is corroborated by
the membership probabilities for V60 and 02461\_8, which are 91\% and 88\% respectively.

The special cases of V9 and B4 deserve attention.  V9 is the binary closest
to the center and its membership probability is 82\%.
However, it looks a very evolved object in the CMD; its period
(3.2~d) and activity (see above) are also more typical for a Main
Sequence star. Therefore, its membership is very doubtful.

The Cudworth membership probability for B4 is only 40\%, but
in the CMDs B4 belongs to a little ``clump" of very blue stars.
This location is in agreement with the results of Liebert et al. (\cite{liebert94})
and therefore B4 is likely a blue extanded horizontal--branch star belonging to NGC~6791.
The star is classified by M02 and M03 (who consider it a non--member) as an
eclipsing binary, but we note that the light curve could also result from a rotational modulation.

\begin{figure}[h]
\centering
   \includegraphics[width=0.79\columnwidth,height=0.79\columnwidth]{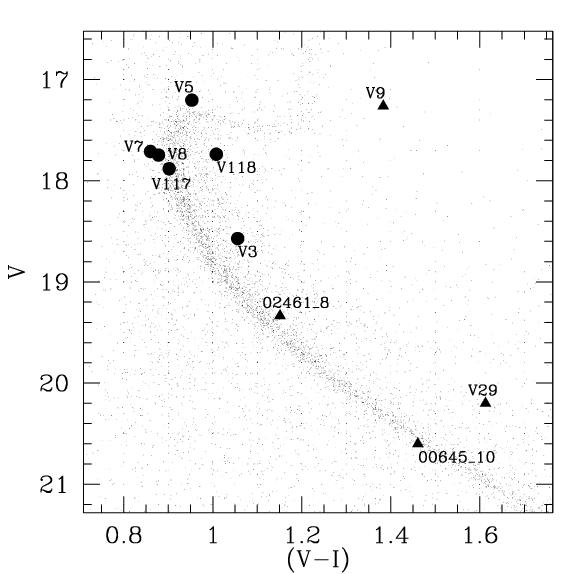}
   \includegraphics[width=0.79\columnwidth,height=0.79\columnwidth]{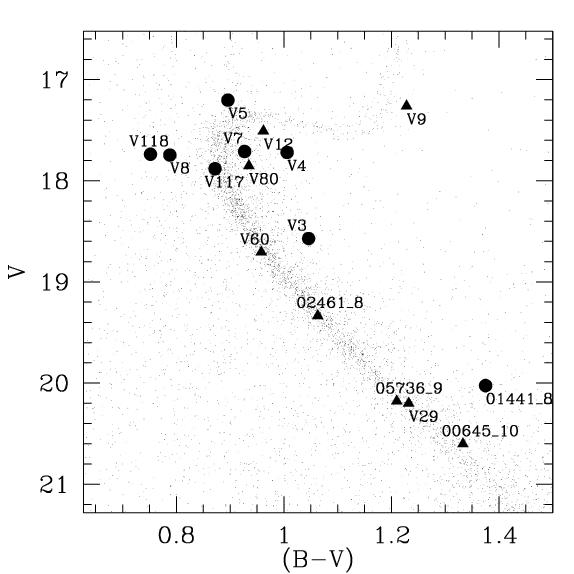}
\caption{\footnotesize
The CMDs for the  binary systems belonging to NGC~6791. Filled circles are contact
binaries (W UMa stars), triangles are detached or semi--detached systems ($\beta$ Per
or $\beta$ Lyr systems).
}
   \label{cmde}
   \end{figure}

Other possible members are: V107, 00331\_3, V109 and
V11, considering that they are within 6\farcm4 radius from the cluster center.
The location in the CMDs of the eclipsing binaries belonging to NGC~6791 is
shown in Fig.~\ref{cmde} (triangles).

\begin{table*}
\begin{flushleft}
\caption{Eclipsing variables with well defined light curves.
EA stands for a $\beta$ Per system, EB  for a  $\beta$ Lyr one.
$V$ is the brightness at maximum
and  $T_0$ is the time of the primary minimum. Also B4, V60 and 02461\_8, whose parameters are
reported in the Appendix, are possible cluster members.
 The last column shows the Cudworth membership
 probability (when available).}

\resizebox{0.99\textwidth}{!}{

\begin{tabular}{llll ccc l ll ll}
\hline \hline
\noalign{\smallskip}
\multicolumn{1}{l}{Star} &
\multicolumn{1}{l}{Type}&
\multicolumn{1}{l}{$\alpha_{2000}$}&
\multicolumn{1}{l}{$\delta_{2000}$}&
\multicolumn{1}{c}{$V$} &
\multicolumn{1}{c}{$<B-V>$} &
\multicolumn{1}{c}{$<V-I>$} &
\multicolumn{1}{l}{Ref.} &
\multicolumn{1}{l}{$T_0$}&
\multicolumn{1}{l}{Period}&
\multicolumn{1}{l}{Ampl.} &
\multicolumn{1}{l}{Notes}  \\
 & & & &[mag] &[mag] &[mag] & &
\multicolumn{1}{l}{[HJD--2452400]}&
\multicolumn{1}{l}{[d]}&
\multicolumn{1}{l}{[mag]} &
\multicolumn{1}{l}{}  \\

\noalign{\smallskip}
\hline
\noalign{\smallskip}

      V119 &     EB &  19.351961 &  37.916328 &   18.15 &    1.13  &    1.33 &    k & 59.879  &    0.30197 &  0.15 &             member ?   \\
       V29 &     EB &  19.354796 &  37.751386 &   20.00 &    1.23  &    1.61 &    k & 69.012  &    0.43662 &  0.22 &         likely memb.   \\
  01558\_5 &     EB &  19.372697 &  37.953469 &   19.15 &          &         &      & 69.028  &    0.52910 &  0.28 &    likely non--memb.   \\
  01393\_1 &     EA &  19.329588 &  37.970392 &   21.23 &          &         &      & 59.326  &    0.58998 &  0.56 &    likely non--memb.   \\
  00331\_3 &     EB &  19.352367 &  37.864391 &   19.73 &    1.20  &    1.36 &    k & 68.815  &     0.7347 &  0.13 &             member ?   \\
       V11 &     EA &  19.342575 &  37.804802 &   19.38 &    0.96  &    1.22 &    k & 67.875  &     0.8833 &  0.48 &             member ?   \\
  05736\_9 &     EA &  19.348484 &  37.721855 &   20.20 &    1.21  &         &    k & 68.333  &      1.210 &  0.29 &         likely memb.   \\
       V12 &     EB &  19.345259 &  37.849083 &   17.52 &    0.96  &         &    k & 64.103  &      1.524 &  0.06 &        member (96\%)   \\
 00645\_10 &     EA &  19.354692 &  37.710104 &   20.60 &    1.33  &    1.46 &    k & 60.893  &      1.451 &  0.20 &         likely memb.   \\
        V9 &     EB &  19.346634 &  37.777035 &   17.15 &    1.23  &    1.38 &    k & 63.873: &        3.2 & $>$0.2 &        member (82\%) \\
      V107 &     EA &  19.355068 &  37.761553 &   17.97 &    0.93  &    1.00 &    k & 64.433  &       3.27 &  0.24 &             member ?   \\
      V109 &     EA &  19.342716 &  37.793961 &   20.73 &    1.46  &    1.60 &    k & 69.021  &       3.70 &  0.86 &             member ?   \\
\noalign{\smallskip}
\hline \hline
\label{tabea}
\end{tabular}
}
\end{flushleft}
\end{table*}

\subsection{Rotational variables}
We found  89  variables whose light curves
are characterized by small amplitude (usually less than 0.10~mag) and
continuous variability.  It is difficult to ascribe such
variability to contact binaries
undergoing grazing eclipses, since they should be less numerous than those having
partial eclipses, since grazing eclipses occur only for a particular orientation
of the orbital plane.
Our hypothesis is that in most cases this variability results from spots
carried by the
stellar rotation;  under this hypothesis, a large variety of light curves can be produced.
Of course, we cannot rule out that  a small fraction of these light curves might be actually generated
by grazing eclipses.

\begin{figure}[h]
\begin{center}
\includegraphics[width=0.30\columnwidth,height=0.30\columnwidth]{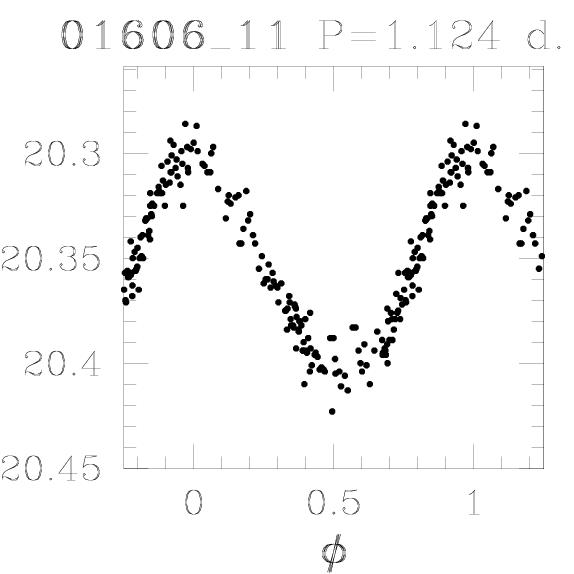}
\includegraphics[width=0.30\columnwidth,height=0.30\columnwidth]{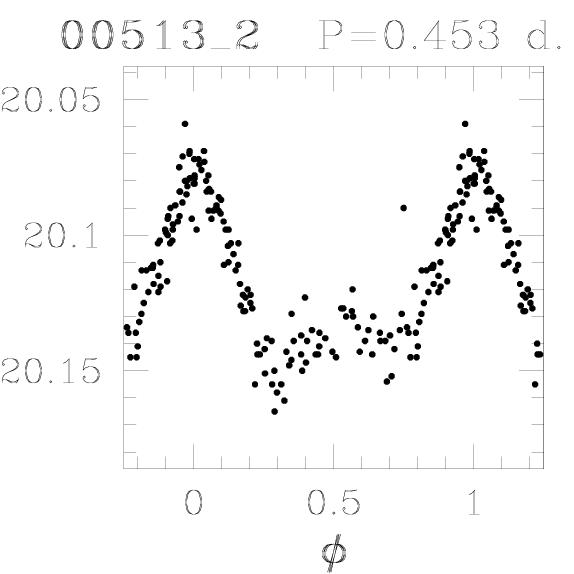}
\includegraphics[width=0.30\columnwidth,height=0.30\columnwidth]{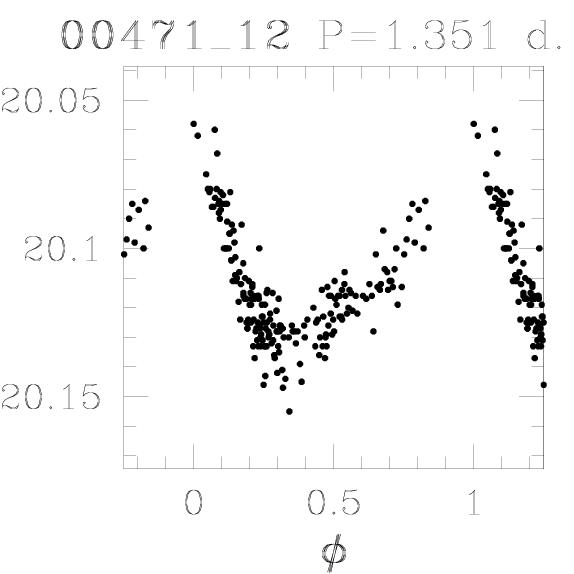}
\includegraphics[width=0.30\columnwidth,height=0.30\columnwidth]{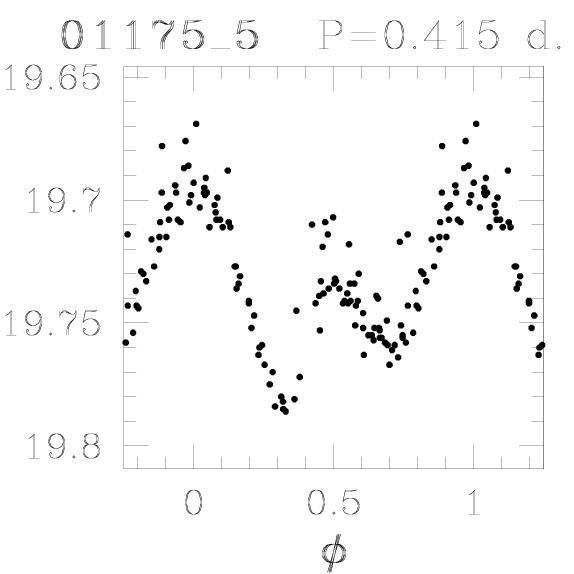}
\includegraphics[width=0.30\columnwidth,height=0.30\columnwidth]{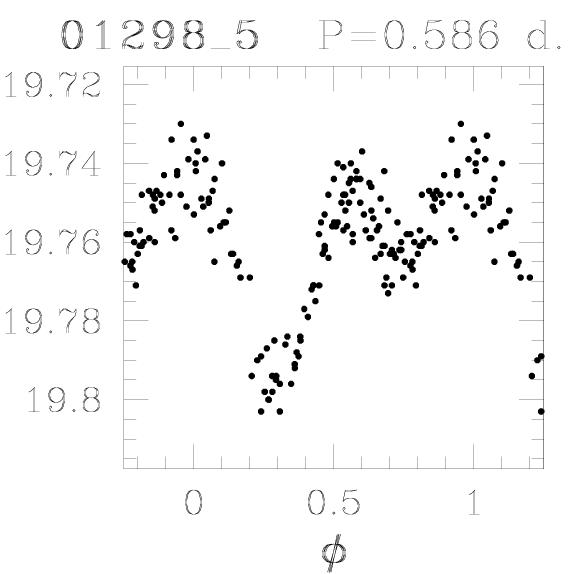}
\includegraphics[width=0.30\columnwidth,height=0.30\columnwidth]{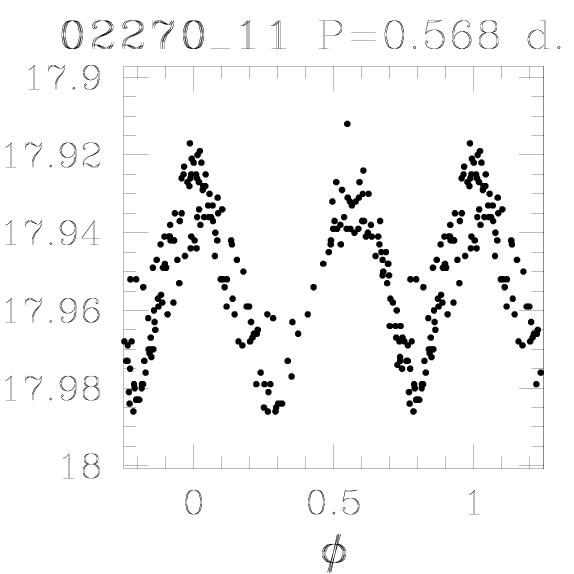}
\caption{\footnotesize The light curves of  a small sample  of rotational variables,
illustrating the growing importance of the second wave.}
\label{roesempio}
\end{center}
\end{figure}

The complete list of the rotational variables and their light curves is given in the Appendix.
Here we discuss some examples.
If the inclination of the rotational axis causes the progressive disappearance of the largest spots,
the light curve displays continuous variation, which could be
sine shaped in the simplest cases (a fraction of the spots is always visible; it can also produce
Cepheid--like variability, as in the  001606\_1 case, Fig.~\ref{roesempio}), or with a standstill
(the hot or cold spots totally disappear; 00513\_2  in Fig.~\ref{roesempio}) or,
more commonly, it can be distorted by other spots besides the
largest ones (00471\_12 in Fig.~\ref{roesempio}).
In cases of very active stars, a secondary wave also occurs (01175\_5 in Fig.~\ref{roesempio}).
Since the second wave often covers less than
half of the period, these rotational variables can be distinguished from eclipsing binaries;
we also note that the amplitude ratio between the first and second waves can be very different.

 Also in the  three cases
in which the full amplitude is larger than 0.10~mag (V2, 02006\_1 and 07483\_9) rotational
effects explain the observed features better than eclipses.
For example, the light curves of V2 ($P$=0.273~d) and 01298\_5 ($P$=0.586~d; see Fig.~\ref{roesempio}) show
typical eclipsing binary  behaviour,
but the amplitudes, the periods, and, mainly, the asymmetries are more typical
of  a rotational effect.
The case of  02270\_11 is different (Fig.~\ref{roesempio}). Its light curve is very similar to
that of a contact binary, but it does not repeat exactly, and unusual scatter is observed through the cycle.
We also note that this non-repetitive behaviour of the light curves, due to the spot activity, is the reason
why several variables stars show residual standard deviations higher than expected.

Our periods for V34, V37 and V38 are approximately half of those given by
M02, since these authors  classified these variables as ellipsoidal ones; the large amplitudes
(0.18, 0.06 and 0.13 mag) are more in favour of a variability resulting from large spots, rather than
the purely geometrical effect of tidally distorted stars. We also note that V37 did not show any flare
activity similar to that reported by M02 during our survey.
We have also revised the classification of V16, considered an eclipsing binary by M02 and  M03.

\begin{figure}[h]
\begin{center}
\includegraphics[width=0.79\columnwidth,height=0.79\columnwidth]{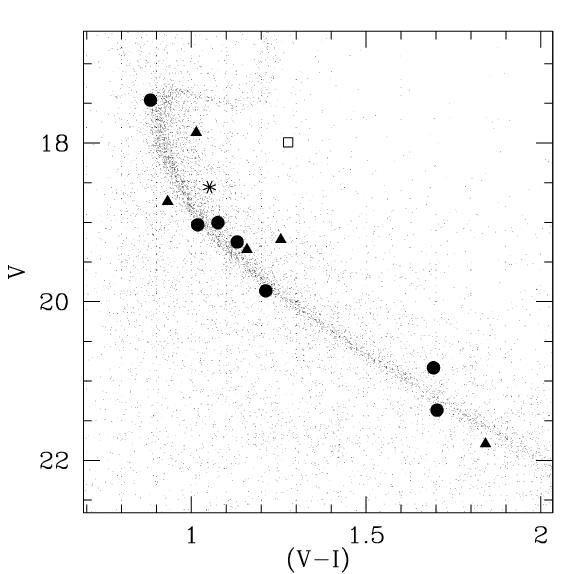}
\includegraphics[width=0.79\columnwidth,height=0.79\columnwidth]{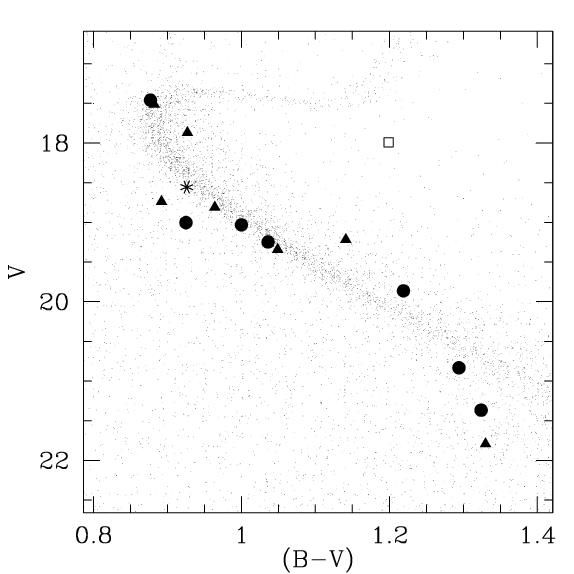}
\caption{\footnotesize $(V-I)-V$ and $(B-V)-V$ diagrams for NGC~6791.
 The rotational variables that we suggest may belong to the cluster
 are indicated with filled circles. {\em Triangles}: stars belonging to the
 cluster according to the Cudworth's membership; {\em starred point}: V14, {\em open
 square}: V17 (see text for details about these stars).}
\label{rocmd}
\end{center}
\end{figure}

 We count 33 rotational variables
in the 10\arcmin--circle (i.e., 0.105~star/arcmin$^2$) centered on the
cluster, while we have 56 variables in the remaining 924--arcmin$^2$ area (i.e., 0.061~star/arcmin$^2$).
We have color indices ($B-V$ and/or $V-I$) for 48 stars;
33 of them have a radial distance less than 10\arcmin\, from the cluster center.
We can confirm the membership for 6 stars having proper motion
membership: V16, V38, V42, V48, V53 and 03079\_9.
We have no photometric indices for V41; however, it is at only 2\arcmin\, from the
cluster center
and its Cudworth probability membership is 77\%. Therefore, we consider V41 a member.
For V14 we have the opposite situation because this star is at 1\arcmin\, from the
cluster center and its positions in the CMDs agree very
well with a membership, but the proper motion measurements rule out that
 it can be a cluster member
(0\%) (see Figure \ref{rocmd}, V14 is displayed as a starred dot).
As mentioned by M03, the positions of V17 in the CMDs are unusual.
Other variables located below the subgiant branch like V17 were found
 in the open cluster M67 (Mathieu et al., \cite{mathieu}) and in the globular cluster
 47 Tuc (Albrow et al., \cite{albrow}). Probably these objects (named ``red
 stragglers'' or ``sub-subgiant branch stars'') are the result of
 some kind of mass exchange between the members of a binary system.

 Putting the rotational variables without proper motion measurements
on the CMDs we could
infer that 8 stars are located on or close to the MS
(represented with filled circles in Fig.~\ref{rocmd}); thus we
suggest that these 8 stars belong to the cluster as well.
Among  the variables at greater distances, for three stars (01149\_2, 01122\_4
and 00513\_2, all located between 11\arcmin\, and 13\arcmin) the membership is
doubtful, since their position in the CMDs is unclear.
The other stars show apparent magnitudes and/or  color indices too discrepant
 to be considered active MS stars belonging to NGC~6791.

When considering the variables without color indices,
only two (V41 and 01874\_2) are at less than 10\arcmin from the cluster center.
We know that V41 is a probable cluster member (membership probability 77\%), but,
 at the moment, we have no valid reason to consider the other star as a member.

Table~\ref{rotbel} lists the rotational variables we suggest as cluster members. The
error bars on the period are $\sim10^{-4}$~d for $P<$1.0~d, $\sim10^{-3}$~d for 1.0$<P<$2.0~d,
$\sim10^{-2}$~d for 2.0$<P<$5.0~d; periods longer than 5.0~d are tentative.
Figure~\ref{rocmd} shows the CMDs with the
rotational variables belonging to the cluster (Tab.~\ref{rotbel}) clearly indicated.
 We rejected as cluster members 16 stars out of 32  located within 10\arcmin\,
from the cluster center; i.e., we considered them to be stars of the Galactic field. We note that the
resulting density of the Galactic field (0.051~star/arcmin$^2$) superimposed
on the cluster is in good agreement with that of the surrounding galactic disk field (0.061~star/arcmin$^2$,
see above), especially considering that Poisson statistics supply uncertainties around $\pm$0.01 on the
density values.

The stellar rotation and the activity level are both expected to be small
for single stars as old as NGC~6791.
Therefore
we suggest that the rotational variables belonging to the cluster are likely short--period binaries,
whose rotational velocity and activity level have been enhanced by the tidal synchronization.

\begin{table*}
\caption{Rotational variables belonging to the cluster. $V$ is the mean brightness value.
$T_0$ indicates the time of the maximum brightness.
The last column shows the membership
 probability (when available) or our photometric membership. The two
uncertain cases (V41 and V14) are also listed (see text).}

\resizebox{0.99\textwidth}{!}{
\begin{tabular}{llll ccc l ll ll}
\hline \hline
\noalign{\smallskip}
\multicolumn{1}{l}{Star} &
\multicolumn{1}{l}{Type}&
\multicolumn{1}{l}{$\alpha_{2000}$}&
\multicolumn{1}{l}{$\delta_{2000}$}&
\multicolumn{1}{c}{$V$} &
\multicolumn{1}{c}{$<B-V>$} &
\multicolumn{1}{c}{$<V-I>$} &
\multicolumn{1}{l}{Ref.} &
\multicolumn{1}{l}{$T_0$}&
\multicolumn{1}{l}{Period}&
\multicolumn{1}{l}{Ampl.} &
\multicolumn{1}{l}{Notes}  \\
 & & & &[mag] & [mag]& [mag] & &
\multicolumn{1}{l}{[HJD--2452400]}&
\multicolumn{1}{l}{[d]}&
\multicolumn{1}{l}{[mag]} &
\multicolumn{1}{l}{}  \\
\noalign{\smallskip}
\hline
\noalign{\smallskip}

  04803\_9 &    RO1 &  19.347698 &  37.796043 &   21.81 &    1.33  &    1.84 &    s & 61.799 &     1.1034 &  0.17 &         member (B06)  \\
       V82 &    RO1 &  19.344366 &  37.793381 &   19.01 &    1.00  &    1.02 &    k & 56.481 &     1.1568 &  0.04 &         likely memb.  \\
  06553\_9 &    RO1 &  19.349134 &  37.672577 &   19.26 &    1.04  &    1.13 &    s & 61.421 &     1.3485 &  0.08 &         likely memb.  \\
  01724\_9 &    RO1 &  19.344957 &  37.785362 &   20.73 &    1.29  &    1.69 &    s & 64.410 &     1.6130 &  0.17 &         likely memb.  \\
       V38 &    RO1 &  19.351021 &  37.768288 &   18.82 &    0.96  &         &    k & 55.630 &       1.96 &  0.13 &        member (92\%)  \\
  03079\_9 &    RO1 &  19.346190 &  37.754753 &   19.23 &    1.14  &    1.26 &    k & 66.630 &      2.640 &  0.07 &        member (93\%)  \\
       V14 &    RO1 &  19.347687 &  37.756874 &   18.58 &    0.93  &    1.05 &    k & 55.933 &       5.45 &  0.05 &    non--member (0\%)  \\
       V48 &    RO1 &  19.352076 &  37.718506 &   17.51 &    0.88  &         &    k & 65.223 &       5.65 &  0.09 &        member (96\%)  \\
       V17 &    RO1 &  19.344135 &  37.817928 &   17.92 &    1.20  &    1.28 &    k & 63.211 &      6.523 &  0.04 &        member (88\%)  \\
       V51 &    RO1 &  19.353382 &  37.748795 &   19.94 &    1.22  &    1.21 &    k & 63.624 &       6.72 &  0.09 &         likely memb.  \\
       V52 &    RO1 &  19.355795 &  37.771935 &   17.49 &    0.88  &    0.88 &    k & 64.345 &       7.06 &  0.03 &         likely memb.  \\
       V53 &    RO1 &  19.350233 &  37.743187 &   18.72 &    0.89  &    0.93 &    k & 69.294 &       7.47 &  0.04 &        member (86\%)  \\
  00436\_3 &    RO2 &  19.352205 &  37.878635 &   18.92 &    0.92  &    1.08 &    k & 60.018 &    0.26601 &  0.04 &         likely memb.  \\
  07483\_9 &    RO2 &  19.349997 &  37.746311 &   21.28 &    1.32  &    1.70 &    s & 60.465 &     0.4375 &  0.17 &         likely memb.  \\
       V41 &    RO2 &  19.347492 &  37.806892 &   19.09 &          &         &      & 60.000 &     0.4798 &  0.07 &        member (77\%)  \\
       V42 &    RO2 &  19.350058 &  37.714867 &   19.51 &    1.05  &    1.16 &    k & 60.323 &     0.5068 &  0.10 &        member (92\%)  \\
       V16 &    RO2 &  19.352108 &  37.802662 &   17.79 &    0.93  &    1.01 &    k & 67.713 &      2.182 &  0.03 &        member (96\%)  \\

\noalign{\smallskip}
\hline \hline
\label{rotbel}
\end{tabular}
}
\end{table*}

\subsection{Long-period variables}
We detected numerous stars having different mean magnitudes on the different nights.
Their behaviours are more diversified than those of the stars we considered
as spurious on the basis of their close similarities.
The resulting power spectra are dominated by terms at very low frequencies, corresponding
to periods often much longer than 10~d. These periods cannot be evaluated in a precise
way, being comparable or, more frequently, longer than our time baseline. Therefore,
we can only argue that these stars are variables, either in a periodic or in an
irregular way. Since we detected many spotted stars, it is quite obvious to think that
most of  these long-period variables are spotted stars having a rotational periods longer than 10~d.
The mean amplitude of these stars is about 0.02 mag, except for 5 stars whose amplitude
exceeds 0.1 mag.

\begin{figure*}[k]
\includegraphics[width=0.79\columnwidth,height=0.79\columnwidth]{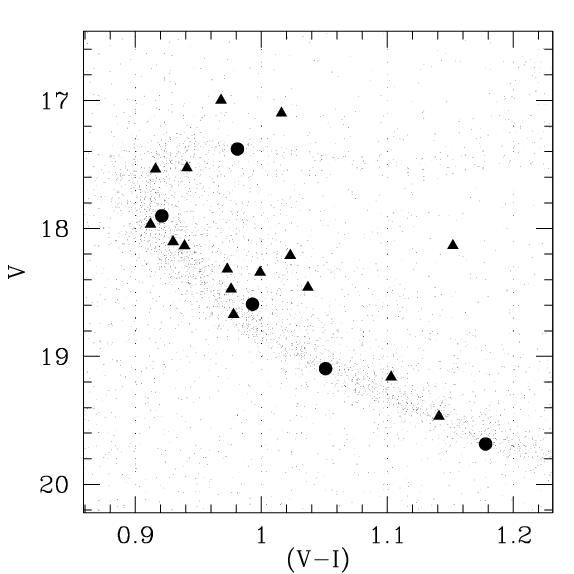}
\includegraphics[width=0.79\columnwidth,height=0.79\columnwidth]{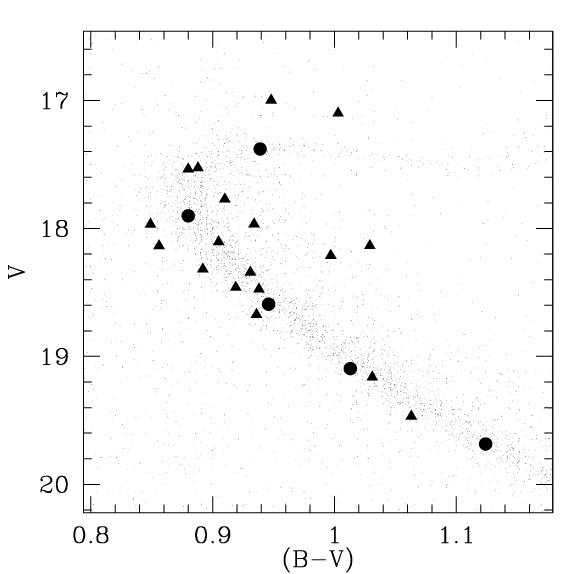}
\caption{\footnotesize $(B-V)-V$ and $(V-I)-V$ diagrams for NGC~6791. {\em Filled circles}:
long--period variables that we suggest belong to the cluster. {\em Filled triangles}:
  long--period variables that belong to the cluster (membership probability
  higher than 76\%.)}
\label{loncmd}
\end{figure*}

\begin{figure*}[h!]
\begin{center}
\includegraphics[width=1.2\columnwidth,height=1.6\columnwidth]{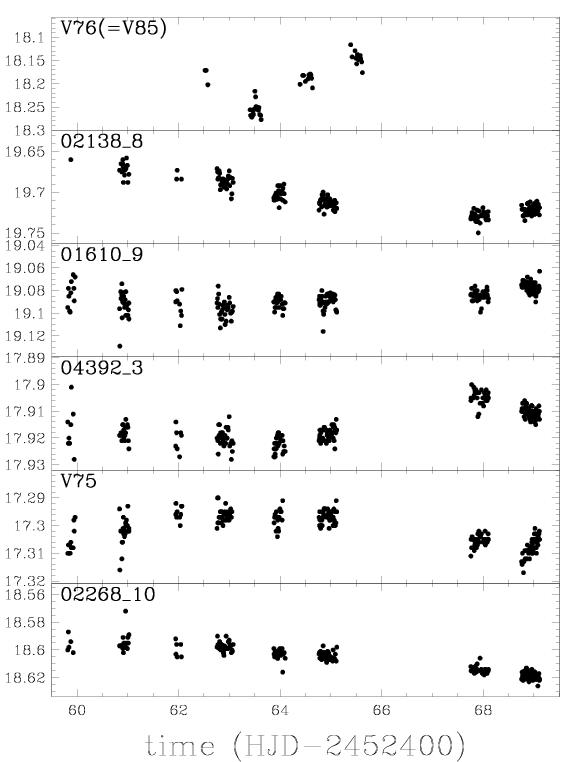}
\caption{\footnotesize Light curve of the suspect ``red straggler''
 V76~$\equiv$~V85 and the five stars that we suggest may
 belong to the cluster. }
\label{lona}
\end{center}
\end{figure*}

 Among the long-period variables, we used the Cudworth probabilities
to establish the membership of 18 stars.
In order to roughly estimate the membership of the remaining
 long-period variables  we checked their locations  in the CMDs,
 in the cases where at least one color is available. We suggest that
 5 stars are likely members of NGC~6791: 02138\_8, 01610\_9, 04392\_3, V75 and
  02268\_10 (see Figure \ref{loncmd}). They lie along the MS or the red-giant branch
and, furthermore, they are all located at
 distances smaller than 8\farcm5, from the cluster center.
Looking at the position of the variable V76~$\equiv$~V85  (memberships: 97\%) in both
 CMDs, we suggest
 that this star could be similar to the ``sub-subgiant branch'' star V17.
In Figure~\ref{lona} we show its light curve and those of the
 5 stars that we suspect to belong to the cluster.
Table~\ref{lptab} lists the long-period variables we suggest as
cluster members; the entire sample is listed in the Appendix.

\begin{table*}
\begin{flushleft}
\caption{Long period variable stars that are likely members 
 of NGC~6791 (ordered by increasing right ascension). The column 9 shows the membership probability (when available) or our photometric membership.}
\begin{tabular}{lllccclllr}
\hline \hline
\noalign{\smallskip}
\multicolumn{1}{l}{Star} &
\multicolumn{1}{l}{$\alpha_{2000}$}&
\multicolumn{1}{l}{$\delta_{2000}$}&
\multicolumn{1}{c}{$V$} &
\multicolumn{1}{c}{$<B-V>$} &
\multicolumn{1}{c}{$<V-I>$} &
\multicolumn{1}{l}{Ref.}&
\multicolumn{1}{l}{Ampl.} &
\multicolumn{1}{l}{Notes}&
\multicolumn{1}{r}{Distance}\\
 & & & [mag] & [mag] & [mag] & & [mag] & & arcmin \\
\noalign{\smallskip}
\hline
\noalign{\smallskip}

  02138\_8  &   19.341864 &  37.749653 &   19.68 &    1.12  &    1.18 &    k &   0.06 &         likely memb.  &   4.6  \\
  02444\_8  &   19.342766 &  37.810604 &   18.34 &    0.93  &    1.00 &    k &   0.02 &        member (78\%)  &   4.4  \\
  00510\_9  &   19.343704 &  37.796719 &   18.67 &    0.94  &    0.98 &    k &   0.03 &        member (83\%)  &   3.4  \\
  01610\_9  &   19.344818 &  37.740486 &   19.09 &    1.01  &    1.05 &    k &   0.02 &         likely memb.  &   3.0  \\
       V94  &   19.345139 &  37.743549 &   17.54 &    0.88  &    0.92 &    k &   0.03 &        member (90\%)  &   2.7  \\
       V95  &   19.345295 &  37.792412 &   19.16 &    1.03  &    1.10 &    k &   0.07 &        member (93\%)  &   2.3  \\
 V56(=V96)  &   19.345908 &  37.763525 &   17.01 &    0.95  &    0.97 &    k &   0.04 &        member (98\%)  &   1.6  \\
  04392\_3  &   19.345982 &  37.870571 &   17.90 &    0.88  &    0.92 &    k &   0.02 &         likely memb.  &   6.1  \\
       V75  &   19.346651 &  37.766308 &   17.38 &    0.94  &    0.98 &    s &   0.01 &         likely memb.  &   1.0  \\
  04133\_9  &   19.347109 &  37.777020 &   18.47 &    0.94  &    0.98 &    k &   0.04 &        member (98\%)  &   0.7  \\
 V76(=V85)  &   19.347192 &  37.764169 &   18.19 &    1.03  &    1.15 &    k &   0.11 &        member (97\%)  &   0.8  \\
       V86  &   19.347258 &  37.808815 &   19.44 &    1.06  &    1.14 &    k &   0.03 &        member (83\%)  &   2.3  \\
       V87  &   19.347996 &  37.749668 &   18.12 &    0.91  &    0.93 &    k &   0.03 &        member (98\%)  &   1.3  \\
  05740\_9  &   19.348534 &  37.808022 &   17.96 &    0.93  &         &    k &   0.03 &        member (95\%)  &   2.2  \\
  06725\_9  &   19.349329 &  37.724495 &   17.77 &    0.91  &         &    k &   0.02 &        member (91\%)  &   3.0  \\
  06796\_9  &   19.349415 &  37.769268 &   18.46 &    0.92  &    1.04 &    k &   0.03 &        member (92\%)  &   1.0  \\
       V90  &   19.349686 &  37.746449 &   18.11 &    0.86  &    0.94 &    k &   0.01 &        member (94\%)  &   1.9  \\
  07680\_9  &   19.350193 &  37.718224 &   18.32 &    0.89  &    0.97 &    k &   0.03 &        member (98\%)  &   3.5  \\
       V31  &   19.350683 &  37.785912 &   17.12 &    1.00  &    1.02 &    k &   0.01 &        member (97\%)  &   2.1  \\
  09376\_9  &   19.351952 &  37.831886 &   18.21 &    1.00  &    1.02 &    k &   0.01 &        member (92\%)  &   4.6  \\
  09611\_9  &   19.352139 &  37.773365 &   17.97 &    0.85  &    0.91 &    k &   0.01 &        member (76\%)  &   2.9  \\
       V58  &   19.354042 &  37.801240 &   17.52 &    0.89  &    0.94 &    k &   0.05 &        member (87\%)  &   4.6  \\
 02268\_10  &   19.359475 &  37.731544 &   18.59 &    0.95  &    0.99 &    k &   0.02 &         likely memb.  &   8.5  \\

\noalign{\smallskip}
\hline \hline
\label{lptab}
\end{tabular}
\end{flushleft}
\end{table*}

\section{Conclusions}
Our wide-field survey of NGC~6791 for the planetary-transit search allowed us to
discover 260 new variable stars. When considering the membership probabilities given
by Cudworth and B06, 13 of them belong to the cluster and one star (09831\_9)
 is not member.
On the basis of the distances from the cluster center and the positions
in the CMDs, we suggest that another 11 stars are
likely members, for 22 stars the membership is doubtful, and 137 stars are likely non-members.
No photometric or kinematic data are available for 76 stars.

The variable star content of the cluster is very similar to that of the surrounding Galactic environment:
in both samples we find rotational variables, contact and eclipsing systems. Contact
binaries and rotational variables belonging to the cluster have the same
characteristics as those located in the surrounding Galactic field. No evidence of
pulsating variables has been found in NGC~6791, but this is not surprising, since
it is a very evolved cluster and stars located in the instability strip or hotter
pulsators have already left the MS.

The discovery of the new cataclysmic variable 06289\_9 in addition to B8 and V15 adds another peculiarity to NGC~6791, making it unusual among
the open clusters.

\label{conclusions}

\begin{acknowledgements}
We are grateful to Kyle Cudworth kindly providing us with preliminary cluster
membership probabilities.
We also acknowledge Prof.~Antonio Bianchini for his suggestions about the characteristics
of the candidate cataclysmic variable and Giovanni Carraro for his useful comments.
We thank the referee, Dr.~J.~Kaluzny, for his detailed report and useful comments.
This work was funded by COFIN 2004
``From stars to planets: accretion, disk evolution and
planet formation'' by MIUR and by  PRIN 2006
``From disk to planetary systems: understanding the origin
and demographics of solar and extrasolar planetary systems''
by INAF.
\end{acknowledgements}

\vfill
\eject
\clearpage

\begin{appendix}
\section{List of identified variables}
\label{list_app}
\begin{figure*}[h]
\begin{center}
\includegraphics[width=1.99\columnwidth,height=2.4\columnwidth]{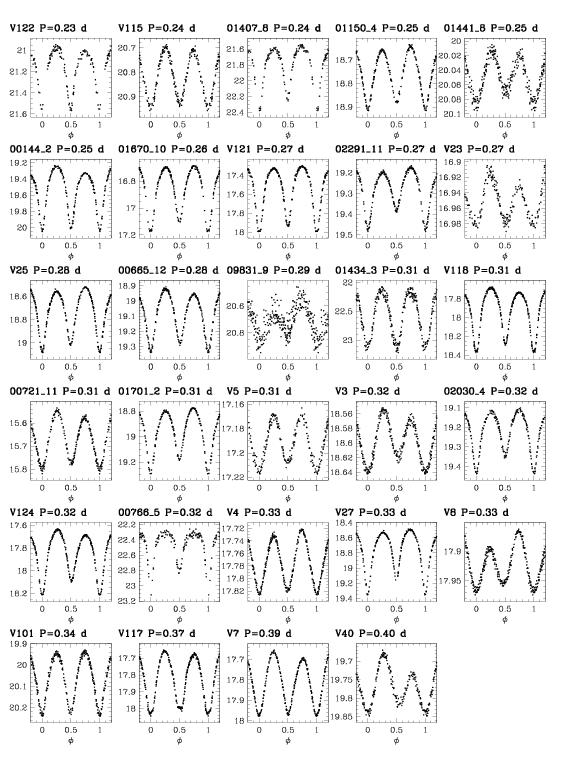}
\caption{\footnotesize Contact variables.}
\label{ew_figure}
\end{center}
\end{figure*}

\begin{figure*}[h]
\begin{center}
\includegraphics[width=1.99\columnwidth,height=2.4\columnwidth]{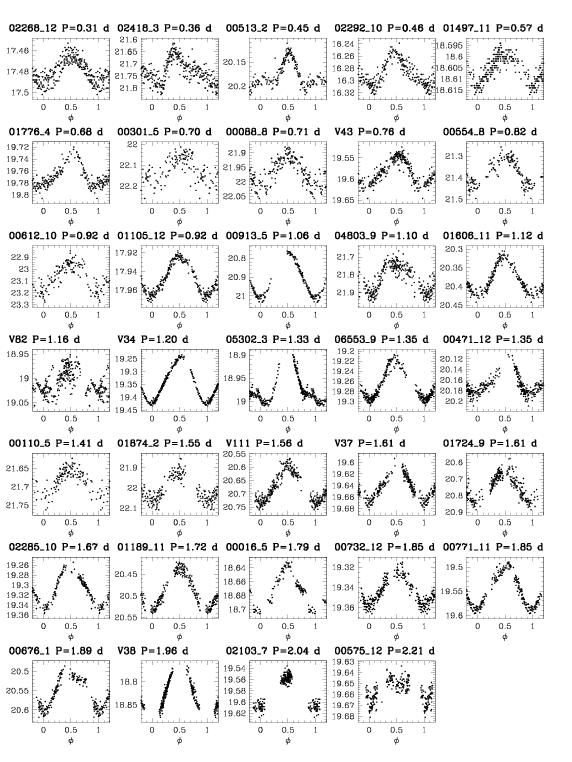}
\caption{\footnotesize Rotational variables with a single--wave light curve,
in some cases very distorted.}
\label{ro1_figure1}
\end{center}
\end{figure*}

\begin{figure*}[h]
\begin{center}
\includegraphics[width=1.99\columnwidth,height=2.4\columnwidth]{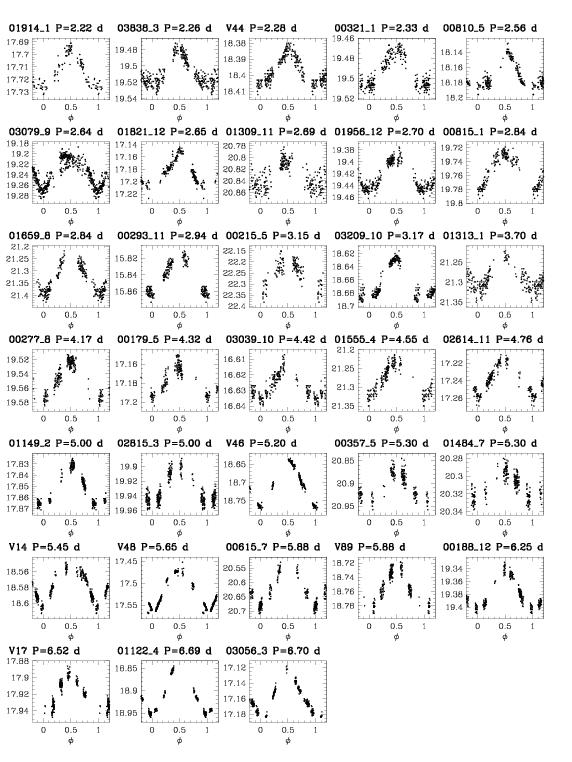}
\caption{\footnotesize Rotational variables with a single--wave light curve,
in some cases very distorted.}
\label{ro1_figure2}
\end{center}
\end{figure*}

\begin{figure*}[h]
\begin{center}
\includegraphics[width=1.99\columnwidth,height=2.4\columnwidth]{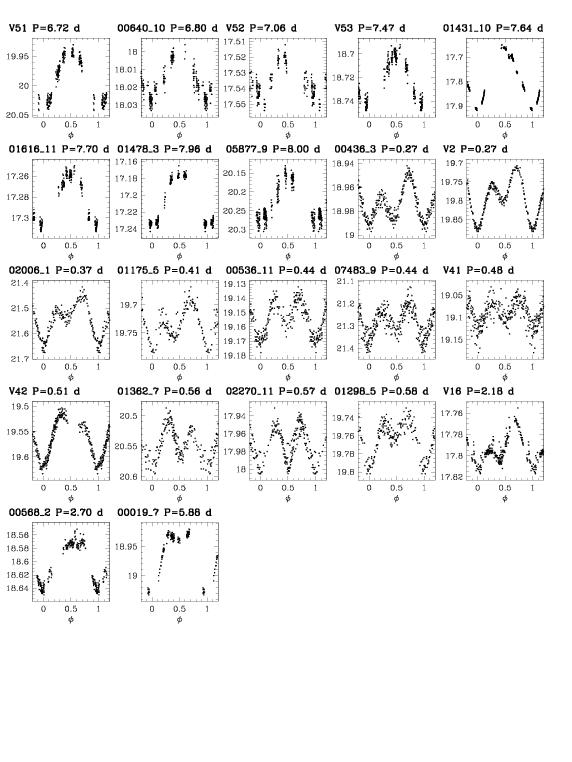}
\caption{\footnotesize Rotational variables with a single--wave (from V51 to 05877\_9)
and a double--wave light curve.}
\label{ro2_figure}
\end{center}
\end{figure*}

This Appendix includes the full list of the identified variables, separated according to our classification:\\
\begin{enumerate}
\item Pulsating, irregular and cataclysmic variables: Table:~\ref{pul_irr_cv_table}.
\item EW--Type stars: Table~\ref{ew_table}, Figure~\ref{ew_figure}.
\item EA/EB--type stars: Table~\ref{eab_table}.
\item Single--wave rotational variables:  Tables~\ref{ro1_table1}, \ref{ro1_table2}, Figures~\ref{ro1_figure1}, \ref{ro1_figure2}.
\item Double--wave rotational variables:  Table~\ref{ro2_table}, Figure~\ref{ro2_figure}.
\item Long--period variables: Table~\ref{lon_table}.
\end{enumerate}

Into the tables, for each star we give the name (a five--digit number followed by the chip number which
the star belongs), coordinates, photometric data
(always the $V$ mag, $B-V$ color when available), informations about the variability
($T_0$, period, amplitude), distance from the center (in arcmin)
 and finally the numerical value of the Cudworth's membership probability (reported in the column 'Memb.').

In most cases, when membership probabilities were not available, in the same column
 the label ``m'' means that we retain the star belonging to the cluster,
while 'm?' and ``nm'' mean ``uncertain membership'' and ``likely non--member'' respectively.
The label ``nd1'' means that no photometric data are available to advance hypothesis about the membership,
 but the star is located nearer than 10\arcmin\, from the center of the cluster.
Finally ``nd2'' means that no photometric data are available and the star is located
 further than 10\arcmin\, from the center; in this case we strongly suggest that the star does not
 belongs to the cluster.

\begin{table*}
\caption{Pulsating, irregular and cataclysmic variables. $V$ is the minimum brightness for
CVs and irregular, the mean brightness for pulsating variables. $T_0$ is the
time of maximum brightness for pulsating stars.}
\label{pul_irr_cv_table}

\begin{flushleft}
\resizebox{0.99\textwidth}{!}{

\begin{tabular}{ll ll ccc l ll llr}
\hline
\noalign{\smallskip}
\multicolumn{1}{l}{Star} &
\multicolumn{1}{l}{Type} &
\multicolumn{1}{l}{$\alpha_{2000}$}&
\multicolumn{1}{l}{$\delta_{2000}$}&
\multicolumn{1}{c}{$V$} &
\multicolumn{1}{c}{$<B-V>$} &
\multicolumn{1}{c}{$<V-I>$} &
\multicolumn{1}{l}{Ref.} &
\multicolumn{1}{l}{$T_0$}&
\multicolumn{1}{l}{Period} &
\multicolumn{1}{l}{Ampl.} &
\multicolumn{1}{l}{Memb.} &
\multicolumn{1}{r}{Distance}\\
 & & & &[mag] &[mag]& [mag] & &
\multicolumn{1}{l}{[HJD--2452400]} &
\multicolumn{1}{l}{[d]} &
\multicolumn{1}{r}{[mag]} & & [arcmin]\\

\noalign{\smallskip}
\hline

\noalign{\smallskip}

      V123 &   HADS &  19.362064 &  37.666034 &   17.08 &    0.45  &         &    k & 59.559 &    0.06026 &  0.14 &          nm  &  11.8  \\
 01497\_12 &   HADS &  19.379083 &  37.812419 &   16.06 &          &         &      & 59.528 &    0.07227 &  0.40 &         nd2  &  22.2  \\
  00311\_7 &  SXPhe &  19.324628 &  37.716768 &   23.17 &          &         &      & 59.605 &    0.10443 &  0.10 &         nd2  &  17.0  \\
 00224\_10 &  SXPhe &  19.353639 &  37.710163 &   21.72 &    0.71  &    1.06 &    s & 59.801 &    0.12261 &  0.20 &          nm  &   5.4  \\
  03653\_3 &    RRc &  19.347147 &  37.992413 &   17.21 &    0.57  &    0.58 &    k & 59.937 &    0.32654 &  0.39 &          nm  &  13.3  \\
  00345\_1 &   RRab &  19.325082 &  37.964170 &   18.28 &          &         &      & 60.151 &    0.57866 &  0.72 &         nd2  &  20.0  \\
       V92 &    IRR &  19.350754 &  37.766876 &   18.10 &    0.91  &         &    k &        &            &  0.10 &           m  &   1.9  \\
       V83 &    IRR &  19.346220 &  37.737232 &   19.10 &    1.02  &    1.05 &    k &        &            &  0.07 &           m  &   2.4  \\
       V93 &    IRR &  19.351452 &  37.785687 &   18.12 &    0.98  &    1.03 &    s &        &            &  0.04 &           m  &   2.6  \\
  V15(=B7) &     CV &  19.352057 &  37.799019 &   18.26 &    0.20  &         &    k &        &            &  0.06 &          98  &   3.3  \\
        B8 &     CV &  19.343262 &  37.747833 &   20.64 &  --0.23  &    0.78 &    k &        &            &  2.27 &           m  &   3.7  \\
  06289\_9 &     CV &  19.348976 &  37.770355 &   22.80 &    0.25  &    0.88 &    s &        &            &  3.10 &     m (B06)  &   0.7  \\

\noalign{\smallskip}
\hline
\end{tabular}
}
\end{flushleft}
\end{table*}

\begin{table*}
\begin{flushleft}
\caption{Contact binaries; $V$ is the brightness at maximum
and  $T_0$ is the
time of the primary minimum.}
\label{ew_table}
\resizebox{0.99\textwidth}{!}{

\begin{tabular}{ll l ccc ll ll lr}
\hline
\noalign{\smallskip}

\multicolumn{1}{l}{Star} &
\multicolumn{1}{l}{$\alpha_{2000}$}&
\multicolumn{1}{l}{$\delta_{2000}$}&
\multicolumn{1}{c}{$V$} &
\multicolumn{1}{c}{$<B-V>$} &
\multicolumn{1}{c}{$<V-I>$} &
\multicolumn{1}{l}{Ref.} &
\multicolumn{1}{l}{$T_0$}&
\multicolumn{1}{l}{Period} &
\multicolumn{1}{l}{Ampl.} &
\multicolumn{1}{l}{Memb.} &
\multicolumn{1}{r}{Distance}\\
 & & & [mag] & [mag] & [mag] & & [HJD--2452400] & [d] & [mag] & & [arcmin]\\
\noalign{\smallskip}
\hline
\noalign{\smallskip}

      V122  &   19.360729 &  37.641436 &   20.92 &          &         &      & 59.614  &    0.22883 &  0.58 &       nd2  &  11.9  \\
      V115  &   19.330646 &  37.975902 &   20.70 &          &         &      & 59.473  &    0.23636 &  0.24 &       nd2  &  17.4  \\
  01407\_8  &   19.339256 &  37.701576 &   21.52 &    0.64  &    1.25 &    k & 59.447  &    0.24155 &  0.82 &        nm  &   7.5  \\
  01150\_4  &   19.357227 &  37.962177 &   18.58 &    1.07  &    0.93 &    k & 59.455  &    0.24510 &  0.27 &        nm  &  13.2  \\
  01441\_8  &   19.339422 &  37.778118 &   19.98 &    1.38  &         &    k & 63.073  &    0.24544 &  0.07 &         m  &   6.2  \\
  00144\_2  &   19.334044 &  38.030701 &   19.19 &          &         &      & 59.457  &    0.24780 &  0.77 &       nd2  &  18.5  \\
 01670\_10  &   19.357625 &  37.681442 &   16.64 &    1.14  &    1.24 &    k & 59.729  &    0.25807 &  0.49 &        nm  &   8.7  \\
      V121  &   19.358063 &  37.932346 &   17.24 &    0.81  &    0.84 &    k & 59.619  &    0.26742 &  0.71 &        nm  &  12.0  \\
 02291\_11  &   19.371624 &  37.795464 &   19.15 &          &         &      & 59.582  &    0.26774 &  0.29 &       nd2  &  16.8  \\
       V23  &   19.338614 &  37.787781 &   16.92 &    1.04  &    1.23 &    k & 59.915  &    0.27180 &  0.07 &        nm  &   6.8  \\
       V25  &   19.328426 &  37.713237 &   18.50 &          &         &      & 59.772  &    0.27730 &  0.56 &       nd2  &  14.4  \\
 00665\_12  &   19.375697 &  37.713244 &   18.89 &          &         &      & 59.472  &    0.28369 &  0.41 &       nd2  &  20.0  \\
  09831\_9  &   19.352356 &  37.780907 &   20.55 &    1.09  &    1.19 &    s & 59.662  &    0.29488 &   0.3 &  nm (B06)  &   3.1  \\
  01434\_3  &   19.350643 &  37.985512 &   22.10 &          &         &      & 59.831  &    0.30581 &  0.97 &       nd2  &  13.0  \\
      V118  &   19.347500 &  37.651222 &   17.68 &    0.75  &    1.01 &    s & 59.912  &    0.30623 &  0.70 &         m  &   7.2  \\
 00721\_11  &   19.365737 &  37.713858 &   15.53 &          &         &      & 59.491  &    0.31008 &  0.26 &       nd2  &  13.1  \\
  01701\_2  &   19.341944 &  38.017998 &   18.72 &          &         &      & 59.534  &    0.31201 &  0.51 &       nd2  &  15.4  \\
        V5  &   19.346258 &  37.813354 &   17.19 &    0.90  &    0.95 &    k & 60.221  &    0.31274 &  0.05 &        98  &   2.8  \\
        V3  &   19.354380 &  37.769349 &   18.51 &    1.05  &    1.06 &    k & 59.798  &    0.31764 &  0.09 &        78  &   4.5  \\
  02030\_4  &   19.353487 &  38.009422 &   19.04 &    1.19  &    1.42 &    k & 59.614  &    0.31797 &  0.33 &        nm  &  14.8  \\
      V124  &   19.365150 &  37.681544 &   17.62 &    0.70  &    1.07 &    k & 59.855  &    0.32014 &  0.58 &        nm  &  13.3  \\
  00766\_5  &   19.367585 &  37.943904 &   22.31 &          &         &      & 59.466  &    0.32363 &  0.78 &       nd2  &  17.3  \\
        V4  &   19.348396 &  37.806652 &   17.72 &    1.01  &         &    k & 59.591  &    0.32568 &  0.10 &        98  &   2.1  \\
       V27  &   19.336275 &  37.648720 &   18.47 &    0.82  &    1.29 &    k & 59.763  &    0.33170 &  0.84 &        nm  &  11.2  \\
        V8  &   19.341938 &  37.865810 &   17.81 &    0.79  &    0.88 &    k & 59.896  &    0.33406 &  0.10 &         m  &   7.1  \\
      V101  &   19.351563 &  37.640388 &   19.94 &    0.56  &         &    k & 59.798  &    0.33480 &  0.29 &        nm  &   8.3  \\
      V117  &   19.343433 &  37.665848 &   17.66 &    0.87  &    0.90 &    k & 59.987  &    0.36644 &  0.38 &         m  &   7.2  \\
        V7  &   19.340271 &  37.821892 &   17.63 &    0.93  &    0.86 &    k & 59.820  &    0.39174 &  0.31 &         m  &   6.3  \\
       V40  &   19.327495 &  37.616839 &   19.67 &          &         &      & 60.101  &    0.39750 &  0.16 &       nd2  &  17.3  \\

\noalign{\smallskip}
\hline
\end{tabular}
}
\end{flushleft}
\end{table*}

\begin{table*}
\caption{Eclipsing variables. $V$ is the brightness at maximum
and  $T_0$ is the time of the primary minimum. }
\label{eab_table}

\begin{flushleft}

\resizebox{0.99\textwidth}{!}{

\begin{tabular}{ll llccc lll llrl}
\hline
\noalign{\smallskip}
\multicolumn{1}{l}{Star} &
\multicolumn{1}{l}{Type} &
\multicolumn{1}{l}{$\alpha_{2000}$}&
\multicolumn{1}{l}{$\delta_{2000}$}&
\multicolumn{1}{c}{$V$} &
\multicolumn{1}{c}{$<B-V>$} &
\multicolumn{1}{c}{$<V-I>$} &
\multicolumn{1}{l}{Ref.} &
\multicolumn{1}{l}{$T_0$}&
\multicolumn{1}{l}{Period} &
\multicolumn{1}{l}{Ampl.} &
\multicolumn{1}{l}{Memb.} &
\multicolumn{1}{r}{Distance} &
\multicolumn{1}{l}{Notes}\\
 & & & &[mag] & [mag] & [mag] & & [HJD--2452400] & [d] & [mag] & & [arcmin] & \\
\noalign{\smallskip}
\hline
\noalign{\smallskip}
\noalign{\smallskip}

      V119 &     EB &  19.351961 &  37.916328 &   18.15 &    1.13  &    1.33 &    k & 59.879  &    0.30197 &  0.15 &     m?  &   9.1  & \\
        B4 &  E:    & 19.353589 &  37.764290 &   17.87 &  --0.13  &  --0.15 &    k & 54.788  &    0.39841 &  0.02 &      40  &   4.0  \\
       V29 &     EB &  19.354796 &  37.751386 &   20.00 &    1.23  &    1.61 &    k & 69.012  &    0.43662 &  0.22 &      m  &   4.9 & Light curve distortion  \\
	   &        &            &            &         &          &         &      &         &            &       &         &       & at maximum light \\
  01558\_5 &     EB &  19.372697 &  37.953469 &   19.15 &          &         &      & 69.028  &    0.52910 &  0.28 &    nd2  &  20.6  & \\
  01393\_1 &     EA &  19.329588 &  37.970392 &   21.23 &          &         &      & 59.326  &    0.58998 &  0.56 &    nd2  &  17.7  & \\
  00331\_3 &     EB &  19.352367 &  37.864391 &   19.73 &    1.20  &    1.36 &    k & 68.815  &     0.7347 &  0.13 &     m?  &   6.4  & \\
       V11 &     EA &  19.342575 &  37.804802 &   19.38 &    0.96  &    1.22 &    k & 67.875  &     0.8833 &  0.48 &     m?  &   4.4  & \\
  05736\_9 &     EA &  19.348484 &  37.721855 &   20.20 &    1.21  &         &    k & 68.333  &      1.210 &  0.29 &      m  &   3.0  & \\
       V12 &     EB &  19.345259 &  37.849083 &   17.52 &    0.96  &         &    k & 64.103  &      1.524 &  0.06 &     96  &   5.1  & \\
 00645\_10 &     EA &  19.354692 &  37.710104 &   20.60 &    1.33  &    1.46 &    k & 60.893  &      1.451 &  0.20 &      m  &   6.0  & \\
        V9 &     EB &  19.346634 &  37.777035 &   17.15 &    1.23  &    1.38 &    k & 63.873  &        3.2 &  0.20 &     82  &   1.1  & \\
      V107 &     EA &  19.355068 &  37.761553 &   17.97 &    0.93  &    1.00 &    k & 64.433  &       3.27 &  0.24 &     m?  &   5.0  & Minima at the very  \\
	   &        &            &            &         &          &         &      &         &            &       &         &        &  beginning of the night. \\
      V109 &     EA &  19.342716 &  37.793961 &   20.73 &    1.46  &    1.60 &    k & 69.021  &       3.70 &  0.86 &     m?  &   4.0  & \\
 00219\_11 &     EA &  19.364048 &  37.827507 &   17.94 &    0.92  &    0.87 &    k & 67.913  &            &  0.52 &     nm  &  11.9  & \\
  00663\_4 &     EA &  19.359652 &  37.894062 &   17.49 &    0.79  &    0.83 &    k & 65.003  &            &  0.49 &     nm  &  11.0  & \\
  00671\_2 &     EA &  19.336983 &  37.903919 &   18.46 &    0.70  &    0.91 &    k & 64.968  &            &  0.07 &     nm  &  11.2  & \\
  00828\_5 &     EA &  19.367917 &  37.885330 &   21.28 &          &         &      & 64.979  &            &  0.38 &    nd2  &  15.7  & \\
  00938\_2 &     EA &  19.338285 &  37.861633 &   20.57 &    1.20  &    1.80 &    k & 67.867  &            &  0.60 &     nm  &   8.8  & \\
 00997\_10 &     EA &  19.355692 &  37.799655 &   19.58 &    0.76  &    0.83 &    k & 68.020  &            &  0.20 &     nm  &   5.7  & \\
  01709\_1 &     EA &  19.330890 &  37.932268 &   18.04 &          &         &      & 64.818  &            &  0.32 &    nd2  &  15.5  & \\
 01731\_10 &     EA &  19.357813 &  37.672819 &   19.27 &    0.67  &    0.80 &    k & 64.783  &            &  0.34 &     nm  &   9.1  & Minimum at the very  \\
	   &        &            &            &         &          &         &      &         &            &       &         &        &  beginning of the night. \\
  01780\_8 &     EA &  19.340677 &  37.655354 &   22.19 &          &    2.25 &    k & 68.969  &            &  0.66 &     nm  &   8.7  & \\
  02461\_8 &     EA &  19.342779 &  37.741798 &   19.34 &    1.06  &    1.15 &    k & 60.968  &            &  0.11 &     88  &   4.2  & \\
  01740\_7 &     EA &  19.330623 &  37.638774 &   21.30 &          &         &      & 69.007  &            &  0.33 &    nd2  &  14.8  & Maybe another minimum\\
	   &        &            &            &         &          &         &      &         &            &       &         &        &  at 64.40 \\
 02045\_12 &     EA &  19.381420 &  37.712211 &   17.16 &          &         &      & 68.030  &            &  0.26 &    nd2  &  24.0  & Other minimum at 63.39\\
 02241\_11 &     EA &  19.371472 &  37.827819 &   19.21 &          &         &      & 68.823  &            &  0.53 &    nd2  &  17.0  & Maybe another minimum\\
	   &        &            &            &         &          &         &      &         &            &       &         &        &  at 63.50 \\
  00346\_5 &     EA &  19.365356 &  37.929481 &   20.16 &          &         &      & 65.014  &            &  0.17 &    nd2  &  15.5  & Other minimum at\\
	   &        &            &            &         &          &         &      &         &            &       &         &        &  at 68.410. Short in time. \\
 00631\_12 &     EA &  19.375473 &  37.649140 &   17.92 &          &         &      & 68.753  &            &  0.30 &    nd2  &  20.9  & Minimum at the very  \\
	   &        &            &            &         &          &         &      &         &            &       &         &        &  beginning of the night.\\
 01511\_10 &     EA &  19.357145 &  37.689907 &   19.74 &    1.45  &    1.86 &    k &         &            &  0.50 &     nm  &   8.1  & Two Minima at night extrema.\\
       V60 &     EA &  19.350189 &  37.762493 &   18.68 &    0.96  &         &    k & 67.951  &            &  0.39 &     91  &   1.6  & \\
       V80 &     EA &  19.351799 &  37.791061 &   17.90 &    0.94  &         &    k & 67.607  &      4.631 &  0.10 &     86  &   2.9  & Shallow eclipse ? \\\
      V108 &     EA &  19.352606 &  37.823467 &   21.15 &    1.27  &    1.90 &    k & 69.080  &            &  0.86 &     nm  &   4.5  & \\

\noalign{\smallskip}
\hline
\end{tabular}
}
\end{flushleft}

\end{table*}

\begin{table*}
\begin{flushleft}
\caption{Rotational variables with a single--wave light curve. $V$ is the mean brightness value.
$T_0$ indicates the time of the maximum brightness.}
\label{ro1_table1}

\resizebox{0.99\textwidth}{!}{
\begin{tabular}{ll lccc ll ll lr}
\hline
\noalign{\smallskip}

\multicolumn{1}{l}{Star} &
\multicolumn{1}{l}{$\alpha_{2000}$}&
\multicolumn{1}{l}{$\delta_{2000}$}&
\multicolumn{1}{c}{$V$} &
\multicolumn{1}{c}{$<B-V>$} &
\multicolumn{1}{c}{$<V-I>$} &
\multicolumn{1}{l}{Ref.} &
\multicolumn{1}{l}{$T_0$}&
\multicolumn{1}{l}{Period} &
\multicolumn{1}{l}{Ampl.} &
\multicolumn{1}{l}{Memb.} &
\multicolumn{1}{r}{Distance}\\
 & & &[mag]&[mag] &[mag] & & [HJD--2452400] & [d] & [mag] & & [arcmin] \\

\noalign{\smallskip}
\hline
\noalign{\smallskip}

 02268\_12 &  19.382495 &  37.798839 &   17.42 &          &         &      & 60.323  &    0.30910 &  0.02 &        nd2  &  24.6  \\
  02418\_3 &  19.349085 &  38.016541 &   21.68 &          &         &      & 59.901  &    0.36498 &  0.12 &        nd2  &  14.7  \\
  00513\_2 &  19.336012 &  37.940765 &   20.12 &    1.18  &    1.51 &    k & 60.786  &    0.45272 &  0.08 &         m?  &  13.3  \\
 02292\_10 &  19.359535 &  37.710207 &   16.24 &    1.11  &    1.25 &    k & 61.298  &    0.46371 &  0.06 &         nm  &   9.0  \\
 01497\_11 &  19.368601 &  37.770512 &   18.58 &          &         &      & 61.737  &    0.56508 &  0.01 &        nd2  &  14.6  \\
  01776\_4 &  19.354509 &  37.935883 &   19.70 &    0.99  &    1.16 &    k & 61.154  &    0.68029 &  0.06 &         nm  &  10.9  \\
  00301\_5 &  19.365041 &  37.906284 &   22.10 &          &         &      & 62.833  &    0.70170 &  0.21 &        nd2  &  14.5  \\
  00088\_8 &  19.333437 &  37.744888 &   21.92 &          &         &      & 65.464  &    0.70594 &  0.11 &        nd2  &  10.5  \\
       V43 &  19.344337 &  37.641777 &   19.58 &    1.62  &    2.97 &    k & 55.976  &    0.75759 &  0.08 &         nm  &   8.2  \\
  00554\_8 &  19.335732 &  37.664570 &   21.32 &    1.22  &    1.48 &    k & 65.543  &      0.821 &  0.13 &         nm  &  10.9  \\
 00612\_10 &  19.354604 &  37.729215 &   22.97 &    0.79  &    2.42 &    s & 68.050  &    0.91581 &  0.24 &         m?  &   5.3  \\
 01105\_12 &  19.377359 &  37.737573 &   19.31 &          &         &      & 62.539  &    0.91746 &  0.04 &        nd2  &  21.0  \\
  00913\_5 &  19.368442 &  37.859339 &   20.89 &          &         &      & 62.405  &    1.06386 &  0.24 &        nd2  &  15.4  \\
  04803\_9 &  19.347698 &  37.796043 &   21.81 &    1.33  &    1.84 &    s & 61.799  &     1.1034 &  0.17 &    m (B06)  &   1.5  \\
 01606\_11 &  19.368978 &  37.736585 &   20.36 &          &         &      & 62.390  &    1.12360 &  0.10 &        nd2  &  15.0  \\
       V82 &  19.344366 &  37.793381 &   19.01 &    1.00  &    1.02 &    k & 56.481  &     1.1568 &  0.04 &          m  &   2.9  \\
       V34 &  19.335878 &  37.736183 &   19.30 &    1.16  &    1.41 &    k & 54.955  &    1.20486 &  0.18 &         nm  &   8.9  \\
  05302\_3 &  19.344669 &  37.968491 &   18.90 &          &         &      & 63.224  &     1.3334 &  0.15 &        nd2  &  12.1  \\
  06553\_9 &  19.349134 &  37.672577 &   19.26 &    1.04  &    1.13 &    s & 61.421  &     1.3485 &  0.08 &          m  &   6.0  \\
 00471\_12 &  19.374883 &  37.674950 &   19.35 &          &         &      & 61.900  &     1.3513 &  0.06 &        nd2  &  20.0  \\
  00110\_5 &  19.363887 &  38.001389 &   21.69 &          &         &      & 62.103  &     1.4085 &  0.08 &        nd2  &  17.8  \\
  01874\_2 &  19.342730 &  37.912987 &   21.96 &          &         &      & 64.944  &     1.5503 &  0.14 &        nd1  &   9.3  \\
      V111 &  19.346970 &  37.812141 &   20.67 &    1.46  &    1.66 &    s & 61.673  &     1.5626 &  0.15 &         nm  &   2.5  \\
       V37 &  19.355072 &  37.852001 &   19.58 &    1.65  &    2.48 &    k & 55.721  &     1.6130 &  0.06 &         nm  &   6.9  \\
  01724\_9 &  19.344957 &  37.785362 &   20.73 &    1.29  &    1.69 &    s & 64.410  &     1.6130 &  0.17 &          m  &   2.4  \\
 02285\_10 &  19.359424 &  37.607155 &   19.26 &    1.34  &    1.57 &    k & 63.495  &     1.6668 &  0.08 &         nm  &  12.8  \\
 01189\_11 &  19.367513 &  37.809413 &   20.46 &          &         &      & 64.500  &     1.7242 &  0.10 &        nd2  &  14.0  \\
  00016\_5 &  19.363396 &  37.997513 &   18.68 &    1.54  &    2.52 &    k & 63.157  &     1.7858 &  0.07 &         nm  &  17.4  \\
 00732\_12 &  19.375889 &  37.717569 &   19.39 &          &         &      & 61.992  &      1.852 &  0.04 &        nd2  &  20.1  \\
 00771\_11 &  19.365949 &  37.766208 &   19.52 &          &         &      & 62.492  &      1.852 &  0.09 &        nd2  &  12.7  \\
  00676\_1 &  19.326459 &  37.929277 &   20.54 &          &         &      & 65.481  &      1.887 &  0.11 &        nd2  &  18.0  \\
       V38 &  19.351021 &  37.768288 &   18.82 &    0.96  &         &    k & 55.630  &       1.96 &  0.13 &         92  &   2.1  \\
  02103\_7 &  19.332028 &  37.696796 &   19.57 &    1.65  &         &    k & 60.891  &      2.041 &  0.05 &         nm  &  12.3  \\
 00575\_12 &  19.375449 &  37.792459 &   19.63 &          &         &      & 65.359  &      2.214 &  0.02 &        nd2  &  19.5  \\
  01914\_1 &  19.331830 &  37.878855 &   17.71 &    0.73  &         &    k & 62.358  &      2.222 &  0.03 &         nm  &  13.2  \\
  03838\_3 &  19.346848 &  37.964615 &   19.45 &    0.91  &    1.05 &    k & 64.159  &      2.261 &  0.04 &         nm  &  11.6  \\
       V44 &  19.326970 &  37.694771 &   18.38 &          &         &      & 58.310  &      2.285 &  0.02 &        nd2  &  15.7  \\
  00321\_1 &  19.324957 &  37.905996 &   19.49 &          &         &      & 65.394  &      2.326 &  0.04 &        nd2  &  18.3  \\
  00810\_5 &  19.367793 &  37.865903 &   18.17 &          &         &      & 67.424  &      2.564 &  0.05 &        nd2  &  15.1  \\
  03079\_9 &  19.346190 &  37.754753 &   19.23 &    1.14  &    1.26 &    k & 66.630  &      2.640 &  0.07 &         93  &   1.7  \\
 01821\_12 &  19.380347 &  37.604033 &   20.13 &          &         &      & 70.829  &      2.647 &  0.06 &        nd2  &  25.1  \\
 01309\_11 &  19.367871 &  37.748052 &   20.81 &          &         &      & 61.125  &      2.692 &  0.05 &        nd2  &  14.2  \\
 01956\_12 &  19.380972 &  37.626492 &   21.39 &          &         &      & 63.833  &      2.699 &  0.06 &        nd2  &  25.0  \\
  00815\_1 &  19.327108 &  37.861623 &   19.75 &          &         &      & 63.358  &      2.836 &  0.05 &        nd2  &  15.8  \\
  01659\_8 &  19.340244 &  37.694355 &   21.27 &    1.51  &    1.85 &    k & 64.446  &      2.840 &  0.17 &         nm  &   7.2  \\
 00293\_11 &  19.364252 &  37.703417 &   15.82 &    0.80  &    0.88 &    k & 63.606  &      2.941 &  0.05 &         nm  &  12.2  \\
  00215\_5 &  19.364447 &  37.898928 &   22.26 &          &         &      & 65.340  &       3.15 &  0.14 &        nd2  &  13.9  \\
 03209\_10 &  19.362653 &  37.732028 &   18.61 &    1.54  &    2.57 &    k & 66.016  &       3.17 &  0.06 &         nm  &  10.7  \\
  01313\_1 &  19.329261 &  38.025661 &   21.27 &          &         &      & 62.466  &      3.704 &  0.10 &        nd2  &  20.3  \\
  00277\_8 &  19.334366 &  37.776566 &   19.52 &    1.54  &    2.68 &    k & 60.691  &      4.167 &  0.05 &         nm  &   9.7  \\
  00179\_5 &  19.364283 &  38.002234 &   17.18 &    0.91  &    0.90 &    k & 64.764  &      4.323 &  0.04 &         nm  &  18.0  \\
 03039\_10 &  19.362130 &  37.815850 &   16.58 &    0.79  &    0.87 &    k & 65.634  &      4.423 &  0.02 &         nm  &  10.4  \\
  01555\_4 &  19.355373 &  37.961205 &   21.22 &          &         &      & 69.673  &      4.546 &  0.09 &        nd2  &  12.5  \\
 02614\_11 &  19.372592 &  37.625244 &   17.22 &          &         &      & 70.152  &      4.763 &  0.04 &        nd2  &  19.6  \\
  01149\_2 &  19.339287 &  37.966995 &   17.80 &    0.90  &    0.93 &    k & 67.803  &        5.0 &  0.04 &         m?  &  13.3  \\
  02815\_3 &  19.348427 &  37.967987 &   19.87 &    1.68  &    2.40 &    k & 60.828  &        5.0 &  0.05 &         nm  &  11.8  \\
       V46 &  19.355272 &  37.798869 &   18.65 &    1.18  &    1.36 &    k & 56.886  &        5.2 &  0.14 &         nm  &   5.4  \\
  00357\_5 &  19.365449 &  37.996214 &   20.91 &          &         &      & 63.108  &        5.3 &  0.07 &        nd2  &  18.3  \\
  01484\_7 &  19.329583 &  37.791249 &   20.30 &    1.63  &         &    k & 67.924  &        5.3 &  0.03 &         nm  &  13.2  \\
       V14 &  19.347687 &  37.756874 &   18.58 &    0.93  &    1.05 &    k & 55.933  &       5.45 &  0.05 &          0  &   0.9  \\
       V48 &  19.352076 &  37.718506 &   17.51 &    0.88  &         &    k & 65.223  &       5.65 &  0.09 &         96  &   4.3  \\

\hline
\multicolumn{12}{r}{Continue \dots} \\
\hline

\noalign{\smallskip}
\end{tabular}
}
\end{flushleft}
\end{table*}

\begin{table*}
\begin{flushleft}
\caption{Continue from Table~\ref{ro1_table1}: other rotational variables with a single--wave light curve. $V$ is the mean brightness value.
$T_0$ indicates (one of) the times of maximum brightness.}
\label{ro1_table2}

\resizebox{0.99\textwidth}{!}{
\begin{tabular}{ll l ccc ll ll lr}
\hline
\noalign{\smallskip}

\multicolumn{1}{l}{Star} &
\multicolumn{1}{l}{$\alpha_{2000}$}&
\multicolumn{1}{l}{$\delta_{2000}$}&
\multicolumn{1}{c}{$V$} &
\multicolumn{1}{c}{$<B-V>$} &
\multicolumn{1}{c}{$<V-I>$} &
\multicolumn{1}{l}{Ref.} &
\multicolumn{1}{l}{$T_0$}&
\multicolumn{1}{l}{Period} &
\multicolumn{1}{l}{Ampl.} &
\multicolumn{1}{l}{Memb.} &
\multicolumn{1}{r}{Distance}\\
 & & &[mag]&[mag]&[mag] & & [HJD--2452400] & [d] & [mag] & & [arcmin]\\
\noalign{\smallskip}
\hline
\noalign{\smallskip}

  00615\_7 &  19.325908 &  37.753712 &   20.59 &          &         &      & 59.885  &      5.882 &  0.14 &    nd2  &  15.8  \\
       V89 &  19.349068 &  37.776783 &   18.75 &    0.88  &    1.20 &    s & 63.447  &      5.884 &  0.05 &     nm  &   0.8  \\
 00188\_12 &  19.373720 &  37.632076 &   21.78 &          &         &      & 67.253  &       6.25 &  0.06 &    nd2  &  20.1  \\
       V17 &  19.344135 &  37.817928 &   17.92 &    1.20  &    1.28 &    k & 63.211  &      6.523 &  0.04 &     88  &   3.9  \\
  01122\_4 &  19.357346 &  37.914520 &   18.84 &    1.03  &    1.12 &    k & 65.843  &       6.69 &  0.11 &     m?  &  10.8  \\
  03056\_3 &  19.347979 &  37.869431 &   17.10 &    0.89  &    0.97 &    k & 66.673  &       6.70 &  0.07 &     nm  &   5.9  \\
       V51 &  19.353382 &  37.748795 &   19.94 &    1.22  &    1.21 &    k & 63.624  &       6.72 &  0.09 &      m  &   4.0  \\
 00640\_10 &  19.354585 &  37.604724 &   17.97 &    0.68  &    0.77 &    k & 66.173  &        6.8 &  0.03 &     nm  &  11.0  \\
       V52 &  19.355795 &  37.771935 &   17.49 &    0.88  &    0.88 &    k & 64.345  &       7.06 &  0.03 &      m  &   5.5  \\
       V53 &  19.350233 &  37.743187 &   18.72 &    0.89  &    0.93 &    k & 69.294  &       7.47 &  0.04 &     86  &   2.3  \\
 01431\_10 &  19.357001 &  37.763810 &   17.72 &    1.33  &    1.45 &    k & 67.953  &       7.64 &  0.26 &     nm  &   6.4  \\
 01616\_11 &  19.369060 &  37.767818 &   17.26 &          &         &      & 69.157  &       7.70 &  0.05 &    nd2  &  14.9  \\
  01478\_3 &  19.350523 &  37.862484 &   17.15 &    1.03  &    1.11 &    k & 72.123  &       7.96 &  0.07 &     nm  &   5.7  \\
  05877\_9 &  19.348616 &  37.767616 &   20.22 &    1.32  &    1.49 &    k & 64.272  &        8.0 &  0.12 &     nm  &   0.5  \\

\noalign{\smallskip}
\hline
\end{tabular}
}
\end{flushleft}
\end{table*}

\begin{table*}
\begin{flushleft}
\caption{Rotational variables with a double--wave light curve. $V$ is the mean brightness value.$T_0$ indicates (one of) the times of maximum brightness.}

\label{ro2_table}

\resizebox{0.99\textwidth}{!}{
\begin{tabular}{ll lccc l lll lr}
\hline
\noalign{\smallskip}

\multicolumn{1}{l}{Star} &
\multicolumn{1}{l}{$\alpha_{2000}$}&
\multicolumn{1}{l}{$\delta_{2000}$}&
\multicolumn{1}{c}{$V$} &
\multicolumn{1}{c}{$<B-V>$} &
\multicolumn{1}{c}{$<V-I>$} &
\multicolumn{1}{l}{Ref.} &
\multicolumn{1}{l}{$T_0$}&
\multicolumn{1}{l}{Period} &
\multicolumn{1}{l}{Ampl.} &
\multicolumn{1}{l}{Memb.} &
\multicolumn{1}{r}{Distance}\\
 & & &[mag] & [mag]& [mag] & & [HJD--2452400]  & [d] & [mag] & & [arcmin] \\

\noalign{\smallskip}
\hline
\noalign{\smallskip}

  00436\_3 &  19.352205 &  37.878635 &   18.92 &    0.92  &    1.08 &    k & 60.018  &    0.26601 &  0.04 &      m  &   7.1  \\
        V2 &  19.354875 &  37.766689 &   19.74 &    0.93  &    1.21 &    k & 59.712  &    0.27344 &  0.17 &     nm  &   4.9  \\
  02006\_1 &  19.332258 &  38.043080 &   21.54 &          &         &      & 60.448  &    0.37500 &  0.19 &    nd2  &  19.8  \\
  01175\_5 &  19.370209 &  37.998342 &   19.73 &          &         &      & 59.918  &    0.41481 &  0.09 &    nd2  &  20.8  \\
 00536\_11 &  19.365067 &  37.716529 &   19.13 &          &         &      & 60.460  &    0.43740 &  0.03 &    nd2  &  12.6  \\
  07483\_9 &  19.349997 &  37.746311 &   21.28 &    1.32  &    1.70 &    s & 60.465  &     0.4375 &  0.17 &      m  &   2.1  \\
       V41 &  19.347492 &  37.806892 &   19.09 &          &         &      & 60.000  &     0.4798 &  0.07 &     77  &   2.2  \\
       V42 &  19.350058 &  37.714867 &   19.51 &    1.05  &    1.16 &    k & 60.323  &     0.5068 &  0.10 &     92  &   3.7  \\
  01362\_7 &  19.329000 &  37.662097 &   20.53 &          &         &      & 60.502  &     0.5560 &  0.08 &    nd2  &  15.1  \\
 02270\_11 &  19.371548 &  37.794338 &   17.95 &          &         &      & 60.525  &     0.5679 &  0.05 &    nd2  &  16.8  \\
  01298\_5 &  19.371044 &  38.002255 &   19.76 &          &         &      & 60.402  &     0.5859 &  0.06 &    nd2  &  21.4  \\
       V16 &  19.352108 &  37.802662 &   17.79 &    0.93  &    1.01 &    k & 67.713  &      2.182 &  0.03 &     96  &   3.4  \\
  00568\_2 &  19.336329 &  37.881542 &   18.52 &          &         &      & 60.007  &      2.704 &  0.08 &    nd2  &  10.6  \\
  00019\_7 &  19.323382 &  37.710172 &   18.95 &          &         &      & 65.473  &      5.882 &  0.12 &    nd2  &  17.9  \\

\noalign{\smallskip}
\hline
\end{tabular}
}
\end{flushleft}
\end{table*}

\begin{table*}
\caption{Long period variable stars (ordered by increasing right ascension).}
\label{lon_table}

\begin{center}

\resizebox{0.99\textwidth}{!}{

\begin{tabular}{lll ccc llr|lllcccllr}
\hline
\noalign{\smallskip}

\multicolumn{1}{l}{Star} &
\multicolumn{1}{l}{$\alpha_{2000}$}&
\multicolumn{1}{l}{$\delta_{2000}$}&
\multicolumn{1}{c}{$V$} &
\multicolumn{1}{c}{$<B-V>$} &
\multicolumn{1}{c}{$<V-I>$} &
\multicolumn{1}{l}{Ref.} &
\multicolumn{1}{l}{Memb.}&
\multicolumn{1}{r}{Distance}&
\multicolumn{1}{l}{Star} &
\multicolumn{1}{l}{$\alpha_{2000}$}&
\multicolumn{1}{l}{$\delta_{2000}$}&
\multicolumn{1}{c}{$V$} &
\multicolumn{1}{c}{$<B-V>$} &
\multicolumn{1}{c}{$<V-I>$} &
\multicolumn{1}{l}{Ref.} &
\multicolumn{1}{l}{Memb.} &
\multicolumn{1}{r}{Distance} \\
& & & [mag] & [mag] & [mag] & & & [arcmin] &
& & & [mag] & [mag] & [mag] & & & [arcmin] \\

\noalign{\smallskip}
\hline
\noalign{\smallskip}

  00091\_1 &  19.323698 &  37.943716 &   19.16 &         &         &      &  nd2 & 20.2 &        V90 &  19.349686 &  37.746449 &   18.11 &    0.86 &    0.94 &    k &   94 &  1.9  \\
  00143\_1 &  19.323947 &  37.927971 &   20.11 &         &         &      &  nd2 & 19.5 &   02011\_3 &  19.349693 &  37.958382 &   19.50 &    1.19 &    1.16 &    k &   nm & 11.3  \\
  00250\_7 &  19.324384 &  37.779393 &   18.32 &         &         &      &  nd2 & 16.8 &   01908\_3 &  19.349867 &  37.947132 &   19.29 &    1.05 &    1.09 &    k &   m? & 10.6  \\
  00326\_7 &  19.324675 &  37.719342 &   17.99 &         &         &      &  nd2 & 16.9 &        V91 &  19.350151 &  37.811325 &   18.07 &    0.83 &    0.91 &    k &   m? &  2.8  \\
  00364\_7 &  19.324832 &  37.798606 &   20.17 &         &         &      &  nd2 & 16.6 &   07680\_9 &  19.350193 &  37.718224 &   18.32 &    0.89 &    0.97 &    k &   98 &  3.5  \\
  00377\_1 &  19.325260 &  38.018518 &   18.72 &         &         &      &  nd2 & 21.9 &   07914\_9 &  19.350483 &  37.822048 &   18.64 &    1.12 &    1.27 &    s &   m? &  3.5  \\
  00466\_7 &  19.325296 &  37.809032 &   14.65 &         &         &      &  nd2 & 16.3 &       V100 &  19.350498 &  37.761631 &   17.13 &    1.02 &         &    k &   m? &  1.8  \\
  00435\_1 &  19.325525 &  37.990790 &   16.89 &         &         &      &  nd2 & 20.7 &        V31 &  19.350683 &  37.785912 &   17.12 &    1.00 &    1.02 &    k &   97 &  2.1  \\
  00537\_1 &  19.325956 &  38.059293 &   18.33 &         &         &      &  nd2 & 23.3 &   01344\_3 &  19.350784 &  37.970558 &   18.17 &    1.33 &    1.48 &    k &   nm & 12.1  \\
  00552\_1 &  19.325972 &  37.941670 &   18.10 &         &         &      &  nd2 & 18.7 &        V62 &  19.350847 &  37.731068 &   19.45 &    1.02 &    1.07 &    k &   m? &  3.1  \\
  00651\_7 &  19.326084 &  37.647564 &   16.89 &         &         &      &  nd2 & 17.3 &   01252\_3 &  19.350937 &  37.911869 &   19.85 &    1.55 &    2.66 &    k &   nm &  8.7  \\
  00819\_7 &  19.326732 &  37.764538 &   18.02 &         &         &      &  nd2 & 15.2 &   08747\_9 &  19.351194 &  37.651299 &   21.32 &    1.50 &    1.71 &    s &   nm &  7.6  \\
  00836\_7 &  19.326803 &  37.684309 &   19.78 &         &         &      &  nd2 & 16.0 &   00990\_3 &  19.351418 &  38.014648 &   17.29 &         &         &      &  nd2 & 14.8  \\
  00804\_1 &  19.327102 &  38.039468 &   21.14 &         &         &      &  nd2 & 21.9 &       V110 &  19.351603 &  37.741802 &   17.75 &    0.71 &    0.76 &    k &   25 &  3.1  \\
  00814\_1 &  19.327140 &  38.035436 &   20.78 &         &         &      &  nd2 & 21.7 &   09376\_9 &  19.351952 &  37.831886 &   18.21 &    1.00 &    1.02 &    k &   92 &  4.6  \\
  00973\_1 &  19.327784 &  37.996820 &   19.17 &         &         &      &  nd2 & 19.7 &   09611\_9 &  19.352139 &  37.773365 &   17.97 &    0.85 &    0.91 &    k &   76 &  2.9  \\
  01095\_7 &  19.327905 &  37.726883 &   15.41 &         &         &      &  nd2 & 14.6 &        V66 &  19.352339 &  37.748669 &   15.95 &    1.29 &    1.39 &    k &   m? &  3.3  \\
  01145\_1 &  19.328562 &  38.051636 &   20.12 &         &         &      &  nd2 & 21.8 &   00038\_3 &  19.352835 &  37.851254 &   19.06 &    0.88 &    0.97 &    k &   m? &  5.9  \\
  01382\_7 &  19.329091 &  37.812623 &   18.99 &         &         &      &  nd2 & 13.7 &  00176\_10 &  19.353423 &  37.611797 &   17.90 &    1.39 &    1.59 &    k &   nm & 10.3  \\
  01536\_7 &  19.329780 &  37.640689 &   17.85 &    0.91 &         &    k &   nm & 15.2 &        V58 &  19.354042 &  37.801240 &   17.52 &    0.89 &    0.94 &    k &   87 &  4.6  \\
  01513\_1 &  19.330106 &  37.876259 &   17.18 &    0.70 &         &    k &   nm & 14.2 &  00630\_10 &  19.354660 &  37.738732 &   18.98 &    1.27 &    1.29 &    k &   m? &  5.1  \\
  01746\_7 &  19.330667 &  37.812027 &   18.84 &    1.08 &         &    k &   nm & 12.6 &  00670\_10 &  19.354823 &  37.806999 &   23.18 &   -0.40 &    0.54 &    s &   m? &  5.3  \\
  01829\_7 &  19.330990 &  37.707912 &   15.20 &    1.31 &         &    k &   nm & 12.7 &   01536\_4 &  19.355383 &  37.882149 &   17.21 &    1.18 &    1.29 &    k &   nm &  8.4  \\
  01873\_1 &  19.331671 &  37.912256 &   19.41 &    0.98 &         &    k &   nm & 14.4 &   01415\_4 &  19.355972 &  38.000843 &   17.60 &    1.50 &    1.92 &    k &   nm & 14.9  \\
  02096\_7 &  19.332013 &  37.776243 &   16.79 &    1.05 &         &    k &   nm & 11.4 &  01169\_10 &  19.356078 &  37.634936 &   15.34 &    0.88 &    1.02 &    k &   nm & 10.0  \\
  02126\_7 &  19.332109 &  37.673400 &   19.20 &    1.38 &         &    k &   nm & 12.8 &        V55 &  19.356228 &  37.841698 &   16.12 &    1.48 &    1.88 &    k &   nm &  7.2  \\
  02004\_1 &  19.332252 &  38.055218 &   19.85 &         &         &      &  nd2 & 20.4 &  01232\_10 &  19.356296 &  37.664158 &   19.31 &    1.41 &    1.64 &    k &   nm &  8.7  \\
  02262\_7 &  19.332626 &  37.658640 &   17.99 &    0.77 &         &    k &   nm & 12.9 &  01225\_10 &  19.356349 &  37.770012 &   21.02 &         &    2.91 &    k &   nm &  5.9  \\
      V114 &  19.333338 &  37.812103 &   17.60 &    1.10 &    1.15 &    k &   m? & 10.7 &  01279\_10 &  19.356519 &  37.757845 &   20.83 &    1.00 &    1.22 &    k &   nm &  6.1  \\
  00285\_2 &  19.334742 &  38.038967 &   19.13 &         &         &      &  nd2 & 18.6 &   01210\_4 &  19.356915 &  37.957123 &   17.69 &    0.75 &    0.86 &    k &   nm & 12.8  \\
  00404\_8 &  19.335062 &  37.790737 &   18.68 &    0.69 &         &    k &   nm &  9.3 &  01417\_10 &  19.356934 &  37.756828 &   17.03 &    1.11 &    1.17 &    k &   nm &  6.4  \\
  00386\_2 &  19.335152 &  37.960213 &   21.51 &         &         &      &  nd2 & 14.6 &   01101\_4 &  19.357428 &  37.882973 &   18.25 &    1.02 &    1.11 &    k &   nm &  9.4  \\
  00620\_8 &  19.335981 &  37.656742 &   18.99 &    1.61 &    2.08 &    k &   nm & 11.0 &   01082\_4 &  19.357555 &  37.938122 &   20.49 &    1.24 &    1.10 &    k &   nm & 12.1  \\
  00791\_2 &  19.337513 &  38.014545 &   18.01 &         &         &      &  nd2 & 16.4 &   01070\_4 &  19.357592 &  37.913498 &   17.53 &    1.49 &    2.15 &    k &   nm & 10.9  \\
  00907\_2 &  19.338146 &  38.041290 &   15.62 &         &         &      &  nd2 & 17.6 &  01682\_10 &  19.357705 &  37.729503 &   18.40 &    0.93 &    0.98 &    k &   m? &  7.3  \\
  01212\_8 &  19.338499 &  37.721958 &   18.45 &    0.99 &    1.10 &    s &   nm &  7.4 &  01807\_10 &  19.358104 &  37.721153 &   18.17 &    1.49 &    2.02 &    k &   nm &  7.8  \\
  01222\_8 &  19.338552 &  37.737770 &   18.26 &    1.12 &    1.18 &    k &   nm &  7.1 &   00920\_4 &  19.358349 &  37.962833 &   17.65 &    0.98 &    1.00 &    k &   nm & 13.6  \\
  01126\_2 &  19.339148 &  37.937424 &   16.93 &         &         &      &  nd2 & 11.8 &  01885\_10 &  19.358407 &  37.824782 &   19.23 &    0.83 &    0.88 &    k &   nm &  8.0  \\
  01122\_2 &  19.339149 &  38.031670 &   18.55 &         &         &      &  nd2 & 16.8 &  02145\_10 &  19.359033 &  37.664901 &   16.66 &    1.24 &    1.36 &    k &   nm & 10.1  \\
  01406\_8 &  19.339247 &  37.695103 &   19.03 &    1.22 &    1.37 &    k &   nm &  7.8 &  02165\_10 &  19.359100 &  37.668317 &   19.60 &    1.54 &    2.31 &    k &   nm & 10.0  \\
       V59 &  19.339304 &  37.806061 &   17.96 &    1.30 &    1.42 &    k &   nm &  6.6 &  02250\_10 &  19.359370 &  37.663215 &   18.71 &    0.99 &    1.12 &    k &   nm & 10.4  \\
  01179\_2 &  19.339432 &  37.995293 &   19.66 &         &         &      &  nd2 & 14.7 &   00717\_4 &  19.359394 &  37.895435 &   18.02 &    1.28 &    1.42 &    k &   nm & 11.0  \\
  01622\_8 &  19.340046 &  37.629841 &   18.09 &    1.42 &    1.64 &    k &   nm & 10.2 &   00718\_4 &  19.359436 &  37.961086 &   20.26 &    1.44 &    1.67 &    k &   nm & 13.9  \\
  02022\_8 &  19.341475 &  37.682644 &   18.58 &    1.27 &    1.44 &    k &   nm &  7.1 &  02268\_10 &  19.359475 &  37.731544 &   18.59 &    0.95 &    0.99 &    k &    m &  8.5  \\
  02108\_8 &  19.341743 &  37.645943 &   19.50 &    0.97 &    1.12 &    k &   nm &  8.8 &   00598\_4 &  19.360041 &  37.997711 &   16.73 &    0.90 &    1.05 &    k &   nm & 16.0  \\
  02138\_8 &  19.341864 &  37.749653 &   19.68 &    1.12 &    1.18 &    k &    m &  4.6 &   00569\_4 &  19.360079 &  37.848969 &   19.88 &    1.23 &    1.28 &    k &   nm &  9.7  \\
  02444\_8 &  19.342766 &  37.810604 &   18.34 &    0.93 &    1.00 &    k &   78 &  4.4 &   00538\_4 &  19.360250 &  37.932350 &   17.58 &    0.96 &    0.99 &    k &   nm & 13.0  \\
  06098\_3 &  19.343510 &  38.050224 &   18.50 &         &         &      &  nd2 & 17.0 &   00422\_4 &  19.360813 &  37.962696 &   18.18 &    1.00 &    1.12 &    k &   nm & 14.6  \\
  05998\_3 &  19.343616 &  37.956688 &   19.21 &    1.39 &    1.52 &    k &   nm & 11.5 &   00381\_4 &  19.360947 &  37.864788 &   19.81 &    1.62 &    2.35 &    k &   nm & 10.7  \\
  05951\_3 &  19.343662 &  37.927708 &   18.06 &         &         &      &  nd1 &  9.9 &   00313\_4 &  19.361338 &  37.894558 &   18.56 &    0.72 &    0.79 &    k &   nm & 12.0  \\
  05938\_3 &  19.343676 &  37.927959 &   18.19 &         &         &      &  nd1 &  9.9 &  02898\_10 &  19.361434 &  37.604020 &   22.56 &         &         &      &  nd2 & 13.9  \\
  00510\_9 &  19.343704 &  37.796719 &   18.67 &    0.94 &    0.98 &    k &   83 &  3.4 &  02924\_10 &  19.361641 &  37.711676 &   18.04 &    1.47 &    1.82 &    k &   nm & 10.3  \\
  05870\_3 &  19.343727 &  37.882679 &   21.96 &         &         &      &  nd1 &  7.3 &  00092\_11 &  19.363431 &  37.702573 &   18.52 &    1.63 &    1.95 &    k &   nm & 11.7  \\
  00835\_9 &  19.344023 &  37.766895 &   20.02 &    1.21 &    1.25 &    k &   m? &  2.9 &   00097\_5 &  19.363811 &  37.998650 &   19.70 &    0.97 &    0.99 &    k &   nm & 17.6  \\
  05677\_3 &  19.344053 &  37.908951 &   18.83 &    1.50 &    1.81 &    k &   nm &  8.7 &  00473\_11 &  19.364836 &  37.670896 &   19.83 &    0.75 &    0.91 &    k &   nm & 13.4  \\
  05524\_3 &  19.344362 &  38.041668 &   20.16 &         &         &      &  nd2 & 16.4 &   00297\_5 &  19.365023 &  37.926561 &   22.15 &         &         &      &  nd2 & 15.2  \\
  01610\_9 &  19.344818 &  37.740486 &   19.09 &    1.01 &    1.05 &    k &    m &  3.0 &  00553\_11 &  19.365157 &  37.749285 &   16.26 &    0.92 &    1.02 &    k &   nm & 12.2  \\
  05014\_3 &  19.345045 &  37.876141 &   16.24 &    1.06 &    1.17 &    k &   m? &  6.6 &   00448\_5 &  19.365911 &  37.852960 &   17.32 &         &         &      &  nd2 & 13.6  \\
       V94 &  19.345139 &  37.743549 &   17.54 &    0.88 &    0.92 &    k &   90 &  2.7 &  00819\_11 &  19.366089 &  37.737134 &   18.77 &         &         &      &  nd2 & 13.0  \\
       V95 &  19.345295 &  37.792412 &   19.16 &    1.03 &    1.10 &    k &   93 &  2.3 &   00521\_5 &  19.366291 &  37.911656 &   17.26 &         &         &      &  nd2 & 15.4  \\
  04768\_3 &  19.345497 &  37.987465 &   19.83 &    1.16 &    1.22 &    k &   m? & 13.1 &   00601\_5 &  19.366800 &  37.992387 &   17.91 &         &         &      &  nd2 & 18.8  \\
  04666\_3 &  19.345663 &  38.023544 &   15.83 &         &         &      &  nd2 & 15.2 &  01021\_11 &  19.366890 &  37.806929 &   17.40 &         &         &      &  nd2 & 13.6  \\
  02716\_9 &  19.345829 &  37.651981 &   17.69 &    1.45 &    1.73 &    k &   nm &  7.4 &   00793\_5 &  19.367687 &  37.917070 &   18.32 &         &         &      &  nd2 & 16.5  \\
 V56(=V96) &  19.345908 &  37.763525 &   17.01 &    0.95 &    0.97 &    k &   98 &  1.6 &  01304\_11 &  19.367749 &  37.637939 &   18.65 &         &         &      &  nd2 & 16.2  \\
  04392\_3 &  19.345982 &  37.870571 &   17.90 &    0.88 &    0.92 &    k &    m &  6.1 &   00922\_5 &  19.368542 &  37.950158 &   18.62 &         &         &      &  nd2 & 18.1  \\
  04368\_3 &  19.346024 &  37.882671 &   19.62 &    1.13 &    1.12 &    k &   m? &  6.8 &  01525\_11 &  19.368695 &  37.773095 &   18.81 &         &         &      &  nd2 & 14.7  \\
  04293\_3 &  19.346100 &  37.838703 &   19.46 &         &         &      &  nd1 &  4.3 &  01695\_11 &  19.369359 &  37.789309 &   17.96 &         &         &      &  nd2 & 15.2  \\
  04298\_3 &  19.346115 &  37.886810 &   18.35 &    0.78 &    0.92 &    k &   nm &  7.0 &  01785\_11 &  19.369770 &  37.820589 &   17.13 &         &         &      &  nd2 & 15.7  \\
  04160\_3 &  19.346313 &  37.864944 &   17.34 &    1.06 &    1.11 &    k &   m? &  5.7 &  02010\_11 &  19.370466 &  37.700018 &   19.73 &         &         &      &  nd2 & 16.5  \\
  03987\_3 &  19.346600 &  37.910511 &   19.17 &    1.04 &    1.10 &    k &   m? &  8.4 &   01245\_5 &  19.370613 &  37.897333 &   18.30 &         &         &      &  nd2 & 17.7  \\
       V75 &  19.346651 &  37.766308 &   17.38 &    0.94 &    0.98 &    s &    m &  1.0 &  02401\_11 &  19.371975 &  37.721131 &   16.62 &         &         &      &  nd2 & 17.3  \\
  03687\_9 &  19.346727 &  37.787651 &   20.13 &    1.07 &    1.45 &    k &   m? &  1.3 &  02419\_11 &  19.371994 &  37.662943 &   17.30 &         &         &      &  nd2 & 18.2  \\
  03859\_3 &  19.346788 &  37.889477 &   20.95 &    1.11 &    1.85 &    k &   nm &  7.1 &  02550\_11 &  19.372438 &  37.653960 &   21.54 &         &         &      &  nd2 & 18.7  \\
  04133\_9 &  19.347109 &  37.777020 &   18.47 &    0.94 &    0.98 &    k &   98 &  0.7 &  00941\_12 &  19.376714 &  37.722466 &   17.77 &         &         &      &  nd2 & 20.6  \\
 V76(=V85) &  19.347192 &  37.764169 &   18.19 &    1.03 &    1.15 &    k &   97 &  0.8 &  01041\_12 &  19.376987 &  37.656970 &   17.89 &         &         &      &  nd2 & 21.7  \\
       V86 &  19.347258 &  37.808815 &   19.44 &    1.06 &    1.14 &    k &   83 &  2.3 &  01038\_12 &  19.377120 &  37.779197 &   20.33 &         &         &      &  nd2 & 20.7  \\
       V87 &  19.347996 &  37.749668 &   18.12 &    0.91 &    0.93 &    k &   98 &  1.3 &  01150\_12 &  19.377457 &  37.669195 &   17.18 &         &         &      &  nd2 & 21.8  \\
  05487\_9 &  19.348234 &  37.677147 &   18.53 &    0.93 &    1.01 &    k &   m? &  5.7 &  01275\_12 &  19.378092 &  37.727700 &   20.16 &         &         &      &  nd2 & 21.5  \\
  05673\_9 &  19.348469 &  37.793930 &   20.61 &    1.29 &    1.62 &    k &   m? &  1.4 &  01807\_12 &  19.380555 &  37.832929 &   18.99 &         &         &      &  nd2 & 23.4  \\
  05740\_9 &  19.348534 &  37.808022 &   17.96 &    0.93 &         &    k &   95 &  2.2 &  01915\_12 &  19.381010 &  37.826766 &   20.20 &         &         &      &  nd2 & 23.7  \\
  06509\_9 &  19.349110 &  37.684639 &   20.79 &    1.37 &    1.79 &    k &   m? &  5.3 &  02001\_12 &  19.381172 &  37.629091 &   20.49 &         &         &      &  nd2 & 25.1  \\
  06532\_9 &  19.349180 &  37.785137 &   18.01 &    0.91 &    0.95 &    s &   m? &  1.1 &  01961\_12 &  19.381188 &  37.780646 &   17.87 &         &         &      &  nd2 & 23.6  \\
  06725\_9 &  19.349329 &  37.724495 &   17.77 &    0.91 &         &    k &   91 &  3.0 &  01968\_12 &  19.381265 &  37.804920 &   18.71 &         &         &      &  nd2 & 23.7  \\
  06796\_9 &  19.349415 &  37.769268 &   18.46 &    0.92 &    1.04 &    k &   92 &  1.0 &  02176\_12 &  19.381966 &  37.690587 &   19.60 &         &         &      &  nd2 & 24.6  \\

\noalign{\smallskip}
\hline
\end{tabular}
}

\end{center}
\end{table*}

\end{appendix}
\end{document}